\documentclass[aps,floatfix,amsmath,nofootinbib,amssymb,superscriptaddress]{revtex4}
\usepackage{overpic}
\usepackage{amssymb}
\usepackage{indentfirst}
\usepackage{feynmf}
\usepackage{slashed}
\usepackage{cases}
\usepackage{color}
\usepackage{multirow}
\usepackage{epstopdf}
\usepackage{graphicx,color,bm}
\usepackage{epstopdf}
\usepackage{amsmath}

\usepackage[colorlinks,
citecolor=blue,
anchorcolor=red,
menucolor=red,
linkcolor=red,
filecolor=red,
runcolor=red,
urlcolor=blue,
frenchlinks=red]{hyperref}

\begin{document}
	
\title{The Sivers asymmetry in charged Kaon and $\Lambda$ hyperon produced SIDIS process at electron ion colliders}
\author{Shuailiang Yang}
\affiliation{School of Physics and Microelectronics, Zhengzhou University, Zhengzhou, Henan 450001, China}
\author{Jianxi Song}
\affiliation{School of Physics and Microelectronics, Zhengzhou University, Zhengzhou, Henan 450001, China}
\author{Xiaoyu Wang}
\email{xiaoyuwang@zzu.edu.cn}
\affiliation{School of Physics and Microelectronics, Zhengzhou University, Zhengzhou, Henan 450001, China}
\author{De-Min Li}
\email{lidm@zzu.edu.cn}
\affiliation{School of Physics and Microelectronics, Zhengzhou University, Zhengzhou, Henan 450001, China}
\author{Zhun Lu}
\email{zhunlu@seu.edu.cn}
\affiliation{School of Physics, Southeast University, Nanjing, Jiangsu 211189, China}

\begin{abstract}
We investigate the single transverse-spin asymmetry with a $\sin (\phi_h-\phi_S)$ modulation in the charged Kaon produced and in $\Lambda$ hyperon produced SIDIS process within the theoretical framework of transverse momentum dependent~(TMD) factorization at the next-to-leading-logarithmic order.
The asymmetry is contributed by the convolution of Sivers function and the unpolarized fragmentation function $D_1$ for the produced hadron.
The parametrization for the proton Qiu-Sterman function, which is closely related to the Sivers function, is adopted to numerically estimate the Sivers asymmetry at the kinematical region of Electron Ion Collider (EIC) and Electron Ion Collider in China (EicC).
The TMD evolution of the TMD parton distribution functions are considered by employing the nonperturbative Sudakov form factor.
It is found that the predicted Sivers asymmetries $A_{UT}^{\sin (\phi_h-\phi_S)}$ as functions of $x$, $z$ and $P_{hT}$ are sizable at the kinematical configurations of both EIC and EicC.
The strange constituent of the produced charged Kaon and $\Lambda$ hyperon in the final state can be a promising probe of the sea quark Sivers function as well as the flavor dependence in the proton target. Therefore, it is important to utilize the future EIC facilities to constrain the sea quark distribution functions as well as the validity of the generalized universality of the Sivers function.
\end{abstract}
	
\maketitle

\section{INTRODUCTION}
Since the measurement by the European Muon Collaboration~\cite{EuropeanMuon:1987isl,EuropeanMuon:1989yki} showed that the spin fraction carried by the internal quarks is much smaller than the spin of the proton, which is contradict to the conventional theoretical prediction that the constituent quark spin contributes the total proton spin, numerous studies have been carried out to explore the nucleon spin structure from both theoretical and experimental aspects. 
Among the spin-related observables, the transverse single spin asymmetries~(TSSAs) can be the key access to the information of transverse momentum structure of nucleon, which is encoded in the transverse momentum dependent parton distribution functions~(TMD PDFs).
In leading twist there are eight TMD PDFs, each one describes a distribution of three-dimensional motion of partons with specified polarization inside the nucleon.
Particularly, the time-reversal odd~(T-odd) Sivers function $f_{1 T}^{\perp}\left(x, p_{T}\right)$~\cite{Sivers:1989cc,Sivers:1990fh} denotes the asymmetric distribution of unpolarized quarks inside a transversely polarized nucleon, which arises from the correlation between the quark transverse momentum and the nucleon transverse spin.
Due to its T-odd property, Sivers function as well as its chiral-odd partner the Boer-Mulders function has been assumed to be forbidden by the naive time-reversal invariance of QCD~\cite{Collins:1992kk}, the very existence of the two T-odd distribution functions was not so obvious.
However, the situation has changed since the calculations in Refs.~\cite{Brodsky:2002cx,Brodsky:2002rv,Boer:2002ju},
which showed that the T-odd distributions can actually survive using spectator model calculations incorporating gluon exchange between the struck quark and the spectator.
In Ref.~\cite{Collins:2002kn}, the time-reversal-invariant argument was reexamined and showed the gauge-link in the operator definition of the correlators guarantee the T-odd distribution functions to be nonzero.
More importantly, the presence of the gauge-link indicates that Sivers function or the Boer-Mulders function has opposite sign between semi-inclusive deeply inelastic scattering~(SIDIS) and Drell-Yan processes~\cite{Brodsky:2002rv,Brodsky:2002cx,Collins:2002kn}
\begin{align}
f_{1 T}^{\perp}\left(x, p_{T}\right)_{[\mathrm{SIDIS}]}=-f_{1 T}^{\perp}\left(x, p_{T}\right)_{[\mathrm{DY}]},
\end{align}
which is a significant prediction by QCD.
The verification of the sign change is one of the most fundamental tests of QCD prediction, and it is also the main pursue of the existing and future Drell-Yan facilities.

The transverse single spin asymmetry can be utilized to extract the information of the Sivers function, and has been intensively investigated in the past two decades from both experimental and theoretical aspects. 
The first non-zero Sivers asymmetry was measured by the HERMES Collaboration at DESY in electroproduction of charge pions off the transversely polarized hydrogen  target~\cite{HERMES:2004mhh}. 
Updated measurements on the Sivers asymmetry in pion produced as well as those in Kaon and $p/\bar{p}$ produced in three-dimensional kinematic bin and enlarged phase space were reported in Refs.~\cite{HERMES:2009lmz,HERMES:2020ifk}.
The COMPASS Collaboration at CERN also measured the Sivers asymmetries in charged hadrons produced SIDIS process through muon beam scattering off the transversely
polarized proton and deuteron targets~\cite{COMPASS:2005csq,COMPASS:2006mkl,COMPASS:2008isr,COMPASS:2010hbb,COMPASS:2012dmt,COMPASS:2016led}.
In addition, the data on the weighted Sivers asymmetry are also released in Ref.~\cite{COMPASS:2018ofp}, allowing for the extractions of the Sivers function
and its first transverse moment.
The Hall A Collaboration at Jefferson Lab presented the measurement of TSSA in charged pion produced SIDIS process with a transversely polarized $^3\mathrm{He}$ target~\cite{JeffersonLabHallA:2011ayy,JeffersonLabHallA:2014yxb}.
Besides the SIDIS process, COMPASS also measured the Sivers asymmetry in Drell-Yan process via $\pi \,N$ collision~\cite{COMPASS:2017jbv}. 
Measurement of TSSAs sensitive to Sivers function in the $W^\pm$ boson produced in proton-proton collisions has also been performed by the STAR experiments at RHIC~\cite{STAR:2015vmv}. 
The data from these measurements have been applied to extract the Sivers function using  parametrizations and phenomenological approaches~\cite{Anselmino:2005ea,Efremov:2004tp,Collins:2005ie,Vogelsang:2005cs,Anselmino:2008sga,Anselmino:2012aa,Bacchetta:2011gx,Echevarria:2014xaa,Anselmino:2016uie,Martin:2017yms,Boglione:2018dqd,Bury:2021sue}.

From the theoretical aspect, intensive studies on the quark Sivers function were performed using the QCD-inspired models, such as the spectator model~\cite{Brodsky:2002cx, Boer:2002ju, Bacchetta:2003rz}, the light-cone quark model~\cite{Lu:2004hu, Pasquini:2010af}, the light-front quark-diquark model~\cite{Maji:2017wwd, Maji:2017zbx}, the non-relativistic constituent quark model~\cite{Courtoy:2008vi}, the MIT bag model~\cite{Yuan:2003wk,Courtoy:2008dn}, and the Holographic QCD~\cite{Lyubovitskij:2022vcl}.
The sea quark Sivers function has been estimated from the light-cone wave function in Refs.~\cite{
Dong:2018wsp,He:2019fzn,Luan:2022fjc}.
However, so far only the valence quark Sivers functions are constrained in the valence region with relatively large uncertainties in the transverse momentum space. One of the reasons is that it is difficult to describe the corresponding physical observables since there are complicated TMD effects. The TMD evolution effects are encoded in the Sudakov-like form factor which also includes details of the non-perturbative QCD dynamics. 
Therefore, this part of the Sudakov-like form factor can not be calculated from the perturbative QCD and is mostly unknown. 
Another reason is that the knowledge of the scale dependence of the TSSAs is very limited since the measurements are mostly performed in the fixed-target experiments with similar hard scales. With the expected high energy and high precision of the Electron Ion Collider~(EIC)~\cite{Accardi:2012qut,AbdulKhalek:2021gbh} and the Electron Ion Collider in China~(EicC)~\cite{Zeng:2022lbo}, the precise knowledge on the TMD distribution functions may be gained, not only for the valence quarks, but also for sea quarks and gluons.
Concerning the sea quark TMDs,
the charged Kaon produced or the Lambda hyperon produced in SIDIS can be recognized as an ideal probe to the sea quark distribution of the proton due to the strange constituent quark inside the kaon and the Lambda hyperon.
Therefore, through the Sivers asymmetry in $K^\pm$ produced and in Lambda produced off transversely polarized nucleon at EIC and EicC, there might be an opportunity to obtain the information of the Sivers distribution function of the sea quark as well as its flavor dependence.

The purpose of this work is to evaluate the Sivers asymmetry in $e p^\uparrow \rightarrow e K^\pm X$ and in $e p^\uparrow \rightarrow e \Lambda X$ at the kinematical region of EIC and EicC.
The theoretical tool adopted in this study is the TMD factorization formalism~\cite{Collins:1981uk,Collins:1984kg,Collins:2011zzd,Ji:2004xq}, which has been widely applied to various high energy processes, such as SIDIS~\cite{Collins:1981uk,Collins:2011zzd,Ji:2004wu,Aybat:2011zv,Collins:2012uy,Echevarria:2012pw}, $e^{+}e^{-}$ annihilation~\cite{Collins:2011zzd,Pitonyak:2013dsu,Boer:2008fr}, Drell-Yan~\cite{Collins:2011zzd,Arnold:2008kf}, and $\mathrm{W}/\mathrm{Z}$ produced in hadron collision~\cite{Collins:1984kg,Collins:2011zzd,Lambertsen:2016wgj}.
In this framework, the differential cross section can be written as the convolution of the well-defined TMD PDFs and/or fragmentation functions~(FFs) at the small transverse momentum region $P_{hT}\ll Q$ as an approximation. The energy dependence of the TMD PDFs and FFs is encoded in the TMD evolution equations, their solutions are usually given in $b$ space, which is conjugate to the transverse momentum space~\cite{Collins:1984kg,Collins:2011zzd} through Fourier transformation.
After solving the TMD evolution equations, the scale dependence of the TMDs may be included in the exponential form of the so-called Sudakov-like form factor~\cite{Collins:1984kg,Collins:2011zzd,Aybat:2011zv,Collins:1999dz}.
The Sudakov-like form factor can be further separated into perturbatively calculable part and the nonperturbative part, the latter one can not be calculated through perturbative theory and may be obtained by fitting experimental data. 
We will consider the TMD evolution effects of the corresponding TMDs to obtain the numerical results for the Sivers asymmetry in charged Kaon $K^\pm$ and $\Lambda$ hyperon produced SIDIS process.

The rest of the paper is organized as follows. In Sec.~\ref{sec:formalisms}, we provide the theoretical framework of Sivers asymmetry $A_{UT}^{\sin (\phi_h-\phi_S)}$ in the charged Kaon produced and $\Lambda$ hyperon produced in SIDIS process within the TMD factorization formalism. 
In Sec.~\ref{sec:number}, we perform the numerical estimation of the Sivers asymmetry at the kinematical region of EIC and EicC. In Sec.~\ref{sec:conciustion}, we summarize the work and discuss the results.

\section{FORMALISM OF THE SIVERS ASYMMETRY IN SIDIS PROCESS}
\label{sec:formalisms}
In this section, we will set up the detailed formalism of the Sivers asymmetry with a modulation of $\sin (\phi_h-\phi_S)$ in the charged Kaon or Lambda produced SIDIS process with an unpolarized electron beam scattered off a transversely polarized proton target
\begin{align}
e(\ell)+p(P)^\uparrow \rightarrow e\left(\ell^{\prime}\right)+K^\pm/\Lambda\left(P_{h}\right)+X,
\end{align}
where $\ell$ and $\ell^\prime$ represent the four-momenta of the incoming and outgoing electrons, $P$ is the four-momentum of the target proton, the up-arrow represents the proton is transversely polarized, $P_{h}$ stands for the four-momentum of the final-state hadron, which can be charged Kaon $K^\pm$ or Lambda hyperon. The four-momentum of the exchanged virtual photon is $q=\ell-\ell^\prime$ and the usual defined energy scale is $Q^2=-q^2$. We denote the masses of the proton target and the final-state hadron by $M$ and $M_h$.
To express the differential cross section as well as the physical observables, we adopt the following Lorentz invariants
\begin{align}
S=(P+\ell)^2\,,\quad
x=\frac {Q^2}{2P \cdot q}\,, \quad
y=\frac{P \cdot q}{P \cdot \ell}\,, \quad
z=\frac{P \cdot P_h}{P \cdot q}\,,
\end{align}
where $S$ is the squared center of mass energy,
$x$ represents the Bjorken variable, $y$ represents the lepton~(quark) energy momentum transferring fraction, and $z$ represents the longitudinal momentum fraction of the final fragmented hadron to the parent quark.

\begin{figure}[h]
\centering
\includegraphics[width=0.7\linewidth]{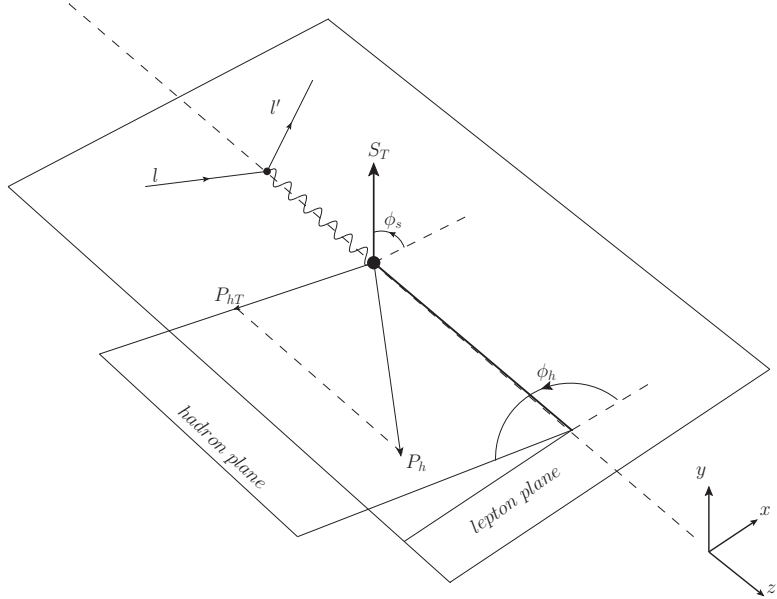}
\caption{ The reference frame in
		SIDIS process.}		
\label{fig:sidis}
\end{figure}

The reference frame applied in our study is shown in Fig.~\ref{fig:sidis}. According to the Trento convention~\cite{Bacchetta:2004jz}, the $z$-axis is defined by the direction of the virtual photon momentum.
The azimuthal angle $\phi_h$ of the outgoing hadron~(charged Kaon or Lambda) is defined by
\begin{equation}
    \cos \phi_h=-\frac{\ell_\mu P_{h\nu} g_\perp^{\mu \nu}}{\sqrt{\ell_T^2 P_{hT}^2}},
    \label{eq:phih}
\end{equation}
with $\ell^\mu_T=g_\perp^{\mu \nu} \ell_\nu$ and $P_{hT}^\mu=g_\perp^{\mu \nu} P_{h\nu}$ being the transverse components of $\ell$ and $P_h$ respect to $z$-axis. The tensor $g_\perp^{\mu \nu}$ is
\begin{equation}
        g_\perp^{\mu \nu}=g^{\mu \nu}-\frac{q^\mu P^\nu + P^\mu q^\nu }{P(1+\gamma^2)}+\frac{\gamma^2}{1+\gamma^2}(\frac{q^\mu q^\nu}{Q^2}-\frac{P^\mu P^\nu}{M^2}),
\end{equation}
with $\gamma=\frac{2Mx}{Q}$.
The azimuthal angle $\phi_S$ of the proton spin vector $S$ is defined by replacing $P_h$ by $S$ in Eq.~(\ref{eq:phih}), and the transverse component of $S$ is $S_{T}^\mu=g_\perp^{\mu \nu} S_{\nu}$ similar to the definition of $P_{hT}$.

Assuming one photon exchange, the model-independent differential cross section can be written as a set of structure functions with the general form as~\cite{Bacchetta:2006tn}
\begin{align}
\frac{d \sigma}{d x d y d z d \phi_{S} d \phi_{h} d P_{h T}^{2}}=\frac{\alpha_{e m}^{2}}{xyQ^{2}}\frac{y}{2(1-\epsilon)}(1+{\gamma^2 \over 2x}) \left\{F_{U U, T}+\left|S_{\perp}\right| \sin \left(\phi_{h}-\phi_{S}\right) F_{U T, T}^{\sin \left(\phi_{h}-\phi_{S}\right)}+\ldots\right\},
\end{align}
where $F_{UU,T}$ stands for the unpolarized structure function, $F_{UT,T}^{\sin(\phi_h-\phi_s)}$ is the transverse spin-dependent structure function contributed by the Sivers function, $\epsilon$ is the ratio of the longitudinal flux and the transverse flux of the photon which has the definition $\epsilon=\frac{1-y-{1\over4}\gamma^2y^2}{1-y+{1\over2}y^2+{1\over4}\gamma^2y^2}$ and the ellipsis denotes other structure functions, which will not be considered in this work.
The three subscripts in the structure functions $F_{XY,Z}$ stand for the polarization of the lepton beam~($X$), the target proton~($Y$) and the virtual photon~($Z$) with U being unpolarized, T being transversely polarized.
The Sivers asymmetry is defined as the ratio of the difference between the spin-dependent differential cross sections and the unpolarized differential cross section
\begin{eqnarray}
A_{U T}^{\sin \left(\phi_{h}-\phi_{s}\right)} \equiv \frac{d \sigma^{\uparrow}-d \sigma^{\downarrow}}{d \sigma^{\uparrow}+d \sigma^{\downarrow}} = \frac{\sigma_{0}\left(x, y, Q^{2}\right)}{\sigma_{0}\left(x, y, Q^{2}\right)} \frac{F_{U T}^{\sin \left(\phi_{h}-\phi_{i}\right)}}{F_{U U}},
\label{eq:asymmetry}
\end{eqnarray}
where $\sigma_{0}=2 \pi \alpha_{\mathrm{em}}^{2}\left[1+(1-y)^{2}\right]$.
The structure functions in Eq.~(\ref{eq:asymmetry}) can be expressed as the convolution of the corresponding TMD PDFs and FFs as~\cite{Bacchetta:2006tn}
\begin{align}
F_{U U, T}&=\mathcal{C}\left[f_{1} D_{1}\right], \label{fuu}\\
F_{U T, T}^{\sin \left(\phi_{h}-\phi_{s}\right)}&=\mathcal{C}\left[-\frac{\boldsymbol{\hat{h}} \cdot \boldsymbol{p}_{T}}{M} f_{1 T}^{\perp} D_{1}\right]. \label{fut}
\end{align}
Here, $f_1(x,\bm{{p}_T})$ is the unpolarized TMD PDF, and $f_{1T}^\perp(x,\bm{p}_T)$ is the Sivers function. $D_1(z,\bm{k}_T)$ is the unpolarized TMD FF, which depends on the longitudinal momentum fraction $z$ and the transverse momentum $\bm{k}_T$ of the final-state quark. $\hat{h}=\frac{\bm{P_{hT}}}{|P_{hT}|}$ is the unit vector along $P_{hT}$. The notation $\mathcal{C}$ represents the convolution among the transverse momenta
\begin{equation}
\label{eq:note_C}
\mathcal {C}\bigl[ \omega f  D \bigr]
= x\sum_q e_q^2 \int d^2 \bm{p}_T  d^2 \bm{k}_T \delta^{(2)}\bigl(\bm{p}_T - \bm{k}_T - \bm{P}_{hT}/z \bigr)\omega(\bm{p}_T,\bm{k}_T)
f^q(x,p_T^2)\,D^q(z,k_T^2).
\end{equation}
Substituting Eq.~(\ref{eq:note_C}) into Eq.~(\ref{fuu}), we can expand the unpolarized structure function $F_{UU}$ as
\begin{align}
F_{U U}\left(Q ; P_{h T}\right) &=\mathcal{C}\left[f_{1} D_{1}\right] \nonumber\\
&=x\sum_{q} e_{q}^{2} \int d^{2} \boldsymbol{p}_{T} d^{2} \boldsymbol{k}_{T} \delta^{(2)}\left(\boldsymbol{p}_{T}-\boldsymbol{k}_{T}-\boldsymbol{P}_{h T} / z\right) f_{1}^{q}\left(x, p_{T}^{2}\right) D_{1}^{q}\left(z, k_{T}^{2}\right) \nonumber\\
&=\frac{x}{z^{2}} \sum_{q} e_{q}^{2} \int d^{2} \boldsymbol{p}_{T} d^{2} \boldsymbol{K}_{T} \delta^{(2)}\left(\boldsymbol{p}_{T}+\boldsymbol{K}_{T} / z-\boldsymbol{P}_{h T} / z\right) f_{1}^{q}\left(x, p_{T}^{2}\right) D_{1}^{q}\left(z, \frac{K_{T}^{2}}{z^{2}}\right) \nonumber\\
&=\frac{x}{z^{2}} \sum_{q} e_{q}^{2} \int d^{2} \boldsymbol{p}_{T} d^{2} \boldsymbol{K}_{T} \int \frac{d^{2} b}{(2 \pi)^{2}} e^{-i\left(p_{T}+\boldsymbol{K}_{T} / z-\boldsymbol{P}_{h T} / z\right) \cdot \boldsymbol{b}} f_{1}^{q}\left(x, p_{T}^{2}\right) D_{1}^{q}\left(z, \frac{K_{T}^{2}}{z^{2}}\right) \nonumber\\
&=\frac{x}{z^{2}} \sum_{q} e_{q}^{2} \int \frac{d^{2} b}{(2 \pi)^{2}} e^{i \boldsymbol{P}_{h T} \cdot \boldsymbol{b} / z } \tilde{f}_{1}^{q/p}\left(x, b\right) \tilde{D}_{1}^{h / q}\left(z, b\right),
\label{fuu_expanded}
\end{align}
where $\boldsymbol{K}_{T}$ represents the transverse momentum of the final state hadron with respect to the fragmentation quark, which has the relation $\bm{K}_T =-z\bm{k}_T$ with $\bm{k_T}$ being the final-state quark transverse momentum respect to $z$ axis. The $\delta$-function Fourier Transformation was performed in the fourth line.
The TMD distribution function $\tilde{f_1}(x,b)$ and TMD fragmentation function $\tilde{D_1}(z,b)$ in the $b$ space can be obtained by performing a Fourier Transformation from momentum space to $b$ space
\begin{align}
\int d^2 \bm{p}_T e^{-i\bm{p}_T \cdot \bm{b}}  f_1^q(x,p_T^2)&=\tilde{f}_1^{q/p}(x,b),\\
\int d^2 \bm{K}_T e^{-i \bm{K}_T/z \cdot \bm{b}} D_1^q(z,K_T^2)&=\tilde{D}_{1}^{h / q}\left(z, b\right),
\label{eq:bspace_und}
\end{align}
hereafter, the term with a tilde denotes it is in the $b$ space.
Similarly, substituting Eq.~(\ref{eq:note_C}) into Eq.~(\ref{fut}), we can obtain expansions for the spin-dependent structure function $F_{UT}^{\sin\left(\phi_h -\phi_s\right)}$ as
\begin{align}
F_{U T, T}^{\sin \left(\phi_{h}-\phi_{S}\right)}\left(Q ; P_{h T}\right)\nonumber &=\mathcal{C}\left[-\frac{\hat{\boldsymbol{h}} \cdot \boldsymbol{p}_{T}}{M} f_{1 T}^{\perp } D_1\right] \nonumber\\
&=x \sum_{q} e_{q}^{2} \int d^{2} \boldsymbol{p}_{T} d^{2} \boldsymbol{k}_{T} \delta^{(2)}\left(\boldsymbol{p}_{T}-\boldsymbol{k}_{T}-\boldsymbol{P}_{h T} / z\right)\left[-\frac{\hat{\boldsymbol{h}} \cdot \boldsymbol{p}_{T}}{M} f_{1 T}^{\perp }\left(x, p_{T}^{2}\right) D_1^q\left(z, k_{T}^{2}\right)\right] \nonumber\\
&=\frac{x}{z^{2}} \sum_{a} e_{q}^{2} \int d^{2} \boldsymbol{p}_{T} d^{2} \boldsymbol{K}_{T} \delta^{(2)}\left(\boldsymbol{p}_{T}+\boldsymbol{K}_{T} / z-\boldsymbol{P}_{h T} / z\right)\left[-\frac{\hat{\boldsymbol{h}} \cdot \boldsymbol{p}_{T}}{M} f_{1 T}^{\perp }\left(x, p_{T}^{2}\right) D_1^q\left(z, \frac{K_{T}^{2}}{z^{2}}\right)\right] \nonumber\\
&=\frac{x}{z^{2}} \sum_{q} e_{q}^{2} \int d^{2} \boldsymbol{p}_{T} d^{2} \boldsymbol{K}_{T} \int \frac{d^{2} b}{(2 \pi)^{2}} e^{-i\left(p_{T}+\boldsymbol{K}_{T} / z-\boldsymbol{P}_{h T} / z\right) \cdot b}\left[-\frac{\hat{\boldsymbol{h}} \cdot \boldsymbol{p}_{T}}{M} f_{1 T}^{\perp }\left(x, p_{T}^{2}\right) D_1^q\left(z, \frac{K_{T}^{2}}{z^{2}}\right)\right] \nonumber\\
&=-\frac{x}{2 z^{2}} \sum_{q} e_{q}^{2} \int \frac{d^{2} b}{(2 \pi)^{2}} e^{i \boldsymbol{P}_{h} \cdot \boldsymbol{b} / z} i \hat{h}_\alpha b^{\alpha} T_{q, F}\left(x, x\right) \tilde{D}_{1}^{h / q}\left(z, b\right).
\label{futt_expanded}
\end{align}
The Sivers function in $b$ space can also be obtained from momentum space $f_1^\perp(x,\bm{p}_T)$ to $b$ space by Fourier Transformation as
\begin{align}
f_{1 T}^{\perp q(\alpha)}(x, b)&=\frac{1}{M} \int d^{2} p_{\perp} e^{-i p_{\perp} \cdot b} p_{\perp}^{\alpha} f_{1 T}^{\perp q}\left(x, p_{\perp}^{2}\right)=\frac{i b^{\alpha}}{2} T_{q, F}\left(x, x\right),
\end{align}
where $T_{q, F}\left(x, x\right)$ is the Qiu-Sterman~(QS) function.
Therefore, the Sivers asymmetry can be rewritten as
\begin{align}
A_{U T}^{\sin \left(\phi_{h}-\phi_{s}\right)}=\frac{\sigma_{0}\left(x, y, Q^{2}\right)}{\sigma_{0}\left(x, y, Q^{2}\right)} \frac{-\frac{x}{2 z^{2}} \sum_{q} e_{q}^{2} \int \frac{d^{2} b}{(2 \pi)^{2}} e^{i \boldsymbol{P}_{h} \cdot \boldsymbol{b} / z} i \hat{\boldsymbol{h}} \cdot b^{\alpha} T_{q, F}\left(x, x\right) \tilde{D}_{1}^{h / q}\left(z, b\right)}{\frac{x}{z^{2}} \sum_{q} e_{q}^{2} \int \frac{d^{2} b}{(2 \pi)^{2}} e^{i \boldsymbol{P}_{h T} \cdot \boldsymbol{b} / z } \tilde{f}_{1}^{q/p}\left(x, b\right) \tilde{D}_{1}^{h / q}\left(z, b\right)}.
\label{AUT1}
\end{align}
One should note that the energy dependence of the TMD structure functions were not encoded in the above formalism, which will be studied in details in the following subsections.

\subsection{TMD evolution effects}
First we set up the basic formalism of the TMD evolution effects for TMD PDFs and FFs, which mainly serves to solve the energy dependence of the TMD PDFs $f_1(x,\bm{{p}_T})$, $f_{1 T}^\perp(x,\bm{p}_T)$ and the TMD FF $D_1(z,\bm{k}_T)$.
Since the complicated convolution among the transverse momenta can be transformed into a simple product after performing the Fourier Transformation, it is convenient to solve the energy dependence in the $b$ space.
		
Particularly, there are two different energy dependencies $\mu$ and $\zeta_F~(\zeta_D)$ of the TMD PDF $\tilde{F}(x,b)$ and the TMD FF $\tilde{D}(z,b)$ in $b$ space according to TMD factorization.
$\mu$ is the renormalization scale related to the corresponding collinear PDFs/FFs, and $\zeta_F~(\zeta_D)$ is the energy scale serving as a cutoff to regularize the light-cone singularity in the operator definition of the TMDs.
The $\mu$ and $\zeta_F~(\zeta_D)$ dependencies are encoded in different TMD evolution equations.
The energy evolution for the $\zeta_F~(\zeta_D)$ dependence is encoded in the Collins-Soper~(CS) equation\cite{Collins:1984kg,Collins:2011zzd,Idilbi:2004vb}
\begin{align}
\frac{\partial \ln \tilde{F}\left(x, b ; \mu, \zeta_{F}\right)}{\partial \ln \sqrt{\zeta_{F}}}=\frac{\partial \ln \tilde{D}\left(z, b ; \mu, \zeta_{D}\right)}{\partial \ln \sqrt{\zeta_{D}}}=\tilde{K}(b ; \mu),
\end{align}
while the $\mu$ dependence is given by the renormalization group equation
\begin{align}
&\frac{d \tilde{K}}{d \ln \mu}=-\gamma_{K}\left(\alpha_{s}(\mu)\right), \\
&\frac{d \ln \tilde{F}\left(x, b ; \mu, \zeta_{F}\right)}{d \ln \mu}=\gamma_{F}\left(\alpha_{s}(\mu) ; \frac{\zeta_{F}^{2}}{\mu^{2}}\right), \\
&\frac{d \ln \tilde{D}\left(z, b ; \mu, \zeta_{D}\right)}{d \ln \mu}=\gamma_{D}\left(\alpha_{s}(\mu) ; \frac{\zeta_{D}^{2}}{\mu^{2}}\right),
\end{align}	
with $\alpha_{s}$ being the running strong coupling at the energy scale $\mu$, $\tilde{K}$ being the CS evolution kernel, and $\gamma_{K},\gamma_{F}$ and $\gamma_{D}$ being the anomalous dimensions.
Hereafter, we will assume $\mu=\sqrt{\zeta_F}=\sqrt{\zeta_D}=Q$, then the TMD PDFs and FFs can be written as $\tilde{F}(x,b;Q)$ and $\tilde{D}(z,b;Q)$ for simplicity.

Solving these TMD evolution equations, one can obtain the solution of the energy dependence for TMD parton distribution functions and the fragmentation functions, of which the solution has the general form as
\begin{align}
\tilde{F}_{q/p}(x,b;Q)=\mathcal{F}\times e^{-S}\times \tilde{F}_{q/p}(x,b;\mu_{B}),\label{eq:f}
\end{align}
\begin{align}
\tilde{D}_{h / q}(z,b;Q)=\mathcal{D}\times e^{-S}\times \tilde{D}_{h / q}(z,b;\mu_{B}),\label{eq:D}
\end{align}
where $\mathcal{F}$ and $\mathcal{D}$ is the factor related to the hard scattering and depend on the factorization schemes, $S$ is the Sudakov-like form factor.
Eq.~(\ref{eq:f}) and Eq.~(\ref{eq:D}) show that the energy evolution of TMD PDFs and TMD fragmentation functions from an initial energy $\mu_{B}$ to another energy $Q$ is encoded in the Sudakov-like form factor $S$ by the exponential form $\mathrm{exp}(-S)$.
	
By performing the reverse Fourier Transformation of the TMDs in $b$ space, the TMDs in momentum space can be obtained, thus it is of great importance to study the $b$ space behavior of the TMDs. 
In the small $b$ region ($b\ll1/\Lambda_{\textrm{QCD}}$), the $b$ dependence of TMDs is perturbative and can be calculated by perturbative QCD.
However, the dependence in large $b$ region turns to be nonperturbative, since the operators are separated by a large distance.
To include the evolution effect in this region, a nonperturbative Sudakov-like form factor $S_{\rm NP}$ is introduced and is usually given in a parameterized form. The parameters of $S_{\rm NP}$ can be determined by analyzing experimental data, given the lack of non-perturbative calculations.
In order to combine the information from both the small $b$ region and the large $b$ region, a matching procedure is applied with a parameter $b_{\mathrm{max}}$ serving as the boundary between the two regions.
Furthermore, one can define a $b$-dependent function $b_\ast$, which has the property $b_\ast\approx b$ in small $b$ region and $b_{\ast}\approx b_{\mathrm{max}}$ in large $b$ region
\begin{align}
b_\ast=\frac{b}{\sqrt{1+b^2/b_{\rm max}^2}} \,  \, , \,  \, b_{\rm max}< 1/\Lambda_{\textrm{QCD}}  \,,
\end{align}
as given in the original CSS prescription~\cite{Collins:1984kg}.
The prescription also allows for a smooth transition from perturbative to nonperturbative regions and avoids the Landau pole singularity in $\alpha_{s}(\mu_{B})$.
The typical value of $b_{\mathrm{max}}$ is chosen around 1.5 GeV$^{-1}$ to guarantee that $b_\ast$ is always in the perturbative region.
With the constraint of $b_\ast$, we can calculate TMDs within a small $b$ region.
	
In the small $b$ region, the TMDs can be expressed as the convolutions of the perturbatively calculable hard coefficients and the corresponding collinear counterparts at fixed energy $\mu_{B}$, which could be the collinear PDFs/FFs or the multiparton correlation functions~\cite{Collins:1981uk,Bacchetta:2013pqa}
\begin{align}
\tilde{F}_{q/p}(x, b ; \mu_{B})= C_{q \leftarrow i} \otimes F_{i/p}(x, \mu_{B}),
\end{align}
\begin{align}
\tilde{D}_{h / q}\left(z, b ; \mu_{B}\right)= \hat{C}_{j \leftarrow q} \otimes D_{h / j}\left(z, \mu_{B}\right),
\end{align}
where $\mu_B=c_0/b_*$ and $c_0=2e^{-\gamma_E}$ and the Euler constant $\gamma_{E} \approx 0.577$~\cite{Collins:1981uk}, the $\otimes$ stands for the convolution in the momentum fraction $x$
\begin{align}
C_{q\leftarrow i}\otimes F_{i/p}(x,\mu_B)& \equiv \sum_{i}\int_{x}^{1} \frac{d \xi}{\xi} C_{q \leftarrow i}\left(\frac{x}{\xi},  \mu_{B} \right) F_{i/p }(\xi, \mu_{B}), \\
\hat C_{j\leftarrow q} \otimes D_{h / j}(z,\mu_B)& \equiv\sum_{j} \int_{z}^{1} \frac{d\xi}{\xi} \hat{C}_{j \leftarrow q}\left(\frac{z}{\xi}, \mu_{B}\right) D_{h / j}\left(\xi, \mu_{B}\right),
\label{eq:otimes}
\end{align}
where $C$ coefficients in the formula has different values in different processes, and its specific value will be given in the subsequent calculation.
In addition, the sum $\sum_{i}$ runs over all parton flavors. Now we can combine all the above information to get the expression for TMD distribution function and the fragmentation function in $b$ space as
\begin{align}
\tilde{F}_{q/p}(x,b;Q)&=\mathcal{F}\times e^{-S}\times  C_{q\leftarrow i}\otimes F_{i/p}(x,\mu_B)\nonumber\\
&=\mathcal{F}\times e^{-S}\times \sum_i \int_{x}^{1} \frac{d \xi}{\xi} C_{q \leftarrow i}\left(\frac{x}{\xi},  \mu_{B} \right) F_{i/p }(\xi, \mu_{B}),\label{eq:f_fixed_evo}\\
\tilde{D}_{h / q}(z,b;Q)&=\mathcal{D}\times e^{-S} \times  \hat{C}_{q\leftarrow i}\otimes D_{h/j}(z,\mu_B)\nonumber\\
&=\mathcal{D}\times e^{-S} \times \sum_j \int_{z}^{1} \frac{d\xi}{\xi} \hat{C}_{j \leftarrow q}\left(\frac{z}{\xi}, \mu_{B}\right) D_{h / j}\left(\xi, \mu_{B}\right).\label{eq:D_fixed_evo}
\end{align}
The Sudakov-like form factor $S$ can be separated into the perturbatively calculable part ${S}_{\rm pert}(Q;b_*)$ and the nonperturbative part $S_{\rm NP}(Q;b)$
\begin{equation}
\label{eq:S}
{S}(Q;b)= {S}_{\rm pert}(Q;b_*)+S_{\rm NP}(Q;b).
\end{equation}
The perturbative part ${S}_{\rm pert}(Q;b_*)$ has a general form and can be expanded as the series of $({\alpha_s \over \pi})$~\cite{Echevarria:2012pw,Echevarria:2014xaa,Kang:2011mr,Aybat:2011ge,Echevarria:2014rua}.
\begin{equation}
\label{eq:Spert}
{S}_{\rm pert}(Q;b_*)=\int^{Q^2}_{\mu_b^2}\frac{d\bar{\mu}^2}{\bar{\mu}^2}\left[A(\alpha_s(\bar{\mu}))
\ln\left(\frac{Q^2}{\bar{\mu}^2}\right)+B(\alpha_s(\bar{\mu}))\right].
\end{equation}
In Eq.~(\ref{eq:Spert}) the coefficients $A$ and $B$ can be expanded as following
\begin{align}
A=\sum_{n=1}^{\infty}A^{(n)}\left(\frac{\alpha_s}{\pi}\right)^n,\\
B=\sum_{n=1}^{\infty}B^{(n)}\left(\frac{\alpha_s}{\pi}\right)^n.
\end{align}
In our calculation we take $A^{(n)}$ up to $A^{(2)}$ and $B^{(n)}$ up to $B^{(1)}$\cite{Collins:1984kg,Aybat:2011zv,Echevarria:2012pw,Kang:2011mr,Landry:2002ix,Qiu:2000ga},
\begin{align}
A^{(1)}&=C_F,\\
A^{(2)}&=\frac{C_F}{2}\left[C_A\left(\frac{67}{18}-\frac{\pi^2}{6}\right)-\frac{10}{9}T_Rn_f\right],\\
B^{(1)}&=-\frac{3}{2}C_F.
\end{align}

For the non-perturbative Sudakov-like form factor $S_{\rm NP}(Q;b)$, it cannot be obtained from perturbation calculation, and it is usually extracted from the experimental data.
Inspired by Refs.~\cite{Landry:2002ix,Konychev:2005iy}, a widely used parametrization (the EIKV parametrization) of $S_{\rm NP}$ for TMD PDFs or fragmentation functions was proposed ~\cite{Bacchetta:2013pqa,Echevarria:2014xaa,Landry:2002ix,Konychev:2005iy,Davies:1984sp,Ellis:1997sc}
\begin{align}
\label{snp}
S_{\mathrm{NP}}^{\mathrm{pdf} / \mathrm{ff}}=b^{2}\left(g_{1}^{\mathrm{pdf} / \mathrm{ff}}+\frac{g_{2}}{2} \ln \frac{Q}{Q_{0}}\right),
\end{align}
where the initial energy is $Q^2_{0}=2.4 ~\mathrm{GeV}^{2}$, and the factor $1/2$ in front of $g_2$ comes from the fact that only one hadron is involved for the parametrization of $S_{\rm NP}^{\rm pdf/ff}$.
The parameter $g_1^{\rm pdf/ff}$ in Eq.~(\ref{snp}) depends on the type of TMDs, which can be regarded as the width of the intrinsic transverse momentum for the relevant TMDs at the initial energy scale $Q_0$~\cite{Aybat:2011zv,Qiu:2000ga,Anselmino:2012aa}.
Assuming a Gaussian form for the dependence for the transverse momentum, one can obtain
\begin{align}
g_1^{\rm pdf} = \frac{\langle p_\perp^2\rangle_{Q_0}}{4},
\qquad
g_1^{\rm ff} = \frac{\langle k_\perp^2\rangle_{Q_0}}{4z_h^2},
\label{eq:g1}
\end{align}
where $\langle p_\perp^2\rangle_{Q_0}$ and $\langle k_\perp^2\rangle_{Q_0}$
represent the averaged intrinsic transverse momenta squared for TMD PDFs and FFs at the initial scale $Q_0$, respectively.
$g_2 =0.16$~\cite{Echevarria:2014xaa} has the same form in all forms of TMDs.
	
As the information on the Sudakov-like form factor for the Kaon fragmentation functions can not be determined, we assume that the evolution of the TMD distribution function and the fragmentation function for producing the $K$ meson from the initial energy scale $\mu$ to another energy scale $Q$ follows the Gaussian form of $g_{2}(b)$ in Eq.~(\ref{snp}), so we can obtain the nonperturbative Sudakov-like form factor for the PDF and FF in the produced of the $K$ mesons as
\begin{align}
S_{\mathrm{NP}}^{\mathrm{pdf}}(Q ; b) &=\frac{g_{2}}{2} \ln \left(\frac{Q}{Q_{0}}\right) b^{2}+g_{1}^{\mathrm{pdf}} b^{2},\label{eq:snppdf}\\
S_{\mathrm{NP}}^{\mathrm{ff}}(Q ; b) &=\frac{g_{2}}{2} \ln \left(\frac{Q}{Q_{0}}\right) b^{2}+g_{1}^{\mathrm{ff}} b^{2}.\label{eq:snpff}
\end{align}
	
Combining all the steps mentioned above together, the scale-dependent TMD PDFs and FFs in $b$ space as functions of $x$ (or $z$), $b$, and $Q$ can be rewritten as
\begin{align}
\tilde{F}_{q/p}(x,b;Q)&=e^{-\frac{1}{2}S_{\mathrm{Pert}}(Q;b_\ast)
-S^{\rm pdf}_{\mathrm{NP}}(Q;b)}\mathcal{F}(\alpha_s(Q))
\sum_i\int_{x}^{1} \frac{d \xi}{\xi} C_{q \leftarrow i}\left(\frac{x}{\xi},  \mu_{B}\right) F_{i/p}(\xi, \mu_{B}),
\label{eq:F_evo1} \\
\tilde{D}_{h/q}(z,b;Q)&=e^{-\frac{1}{2}S_{\mathrm{Pert}}(Q;b_\ast)
-S^{\rm ff}_{\mathrm{NP}}(Q;b)}\mathcal{D}(\alpha_s(Q))
\sum_j \int_{z}^{1} \frac{d\xi}{\xi} \hat{C}_{j \leftarrow q}\left(\frac{z}{\xi}, \mu_{B}\right) D_{h / j}\left(\xi, \mu_{B}\right).
\label{eq:D_evo1}
\end{align}

\subsection{The solution of the unpolarized structure function}
In the following we solve the denominator of the Sivers asymmetry in details, which is the unpolarized structure function $F_{UU}$ in Eq.~(\ref{fuu}).
Since we have expanded $F_{UU}$ in Eq.~(\ref{fuu_expanded}), we will directly follow Eq.~(\ref{fuu_expanded}) to give the complete expression for $F_{UU}$.
We can divide Eq.~(\ref{fuu_expanded}) into three parts to solve the problem: the weighted Bessel function, the distribution function  $\tilde{f}_1^{q/p}(x,b)$ and the fragmentation function $\tilde{D}_1^{h/q}(z,b)$.
The weighted Bessel function can be obtained as
\begin{align}
\frac{x}{z^{2}} \sum_{q} e_{q}^{2}\int \frac{d^{2} b}{(2 \pi)^{2}} e^{i \bm{P_{h T}} \cdot \boldsymbol{b} / z }
&=\frac{x}{z^{2}} \sum_{q} e_{q}^{2}\int_{0}^{\infty}\frac{d b b}{2 \pi} J_{0}\left(\frac{P_{h T}b}{z}\right).
\label{eq:fuucoe}
\end{align}
According to Eq.~(\ref{eq:F_evo1}) and Eq.~(\ref{eq:D_evo1}),  $\tilde{f}_1^q(x,b)$ and $\tilde{D}_1^{h/q}(z,b)$ can be expanded as
\begin{align}
\tilde{f}_1^{q/p}(x,b;Q)&=e^{-\frac{1}{2}S_{\mathrm{Pert}}(Q;b_\ast)
-S^{f_1}_{\mathrm{NP}}(Q;b)}\mathcal{F}(\alpha_s(Q))
\sum_i\int_{x}^{1} \frac{d \xi}{\xi} C_{q \leftarrow i}\left(\frac{x}{\xi},  \mu_{B}\right) f_{1}^{i/p}(\xi, \mu_{B}),\label{eq:ff}\\
\tilde{D}_1^{h/q}(z,b;Q)&=e^{-\frac{1}{2}S_{\mathrm{Pert}}(Q;b_\ast)
-S^{D_1}_{\mathrm{NP}}(Q;b)}\mathcal{D}(\alpha_s(Q))
\sum_j\int_{z}^{1} \frac{d\xi}{\xi} \hat{C}_{j \leftarrow q}\left(\frac{z}{\xi}, \mu_{B}\right) D_1^{h / j}\left(\xi, \mu_{B}\right),\label{eq:dd}
\end{align}
where the Sudakov-like form factor is expressed in Eq.~(\ref{eq:Spert}), Eq.~(\ref{eq:snppdf}) and Eq.~(\ref{eq:snpff}).
The hard scattering coefficients $\mathcal{F}(\alpha_s(Q))$ and $\mathcal{D}(\alpha_s(Q))$ can be set equal to 1.
In addition, the $C$ coefficient in Eq.~(\ref{eq:ff}) and Eq.~(\ref{eq:dd}) can be expressed as~\cite{Kang:2015msa}
\begin{align}
C_{q\gets q'}^{\rm (SIDIS)}(x,\mu_B)&= \delta_{q'q} [\delta(1-x)+\frac{\alpha_s}{\pi}(\frac{C_F}{2}(1-x)  -2C_F\delta(1-x)  )]\; , \label{eq:cf_css}
\\
C_{q\gets g}^{\rm (SIDIS)}(x,\mu_B)&= \frac{\alpha_s}{\pi} {T_R} \, x (1-x)\; ,  \label{eq:cf1_css}
\\
\hat C_{q'\gets q}^{\rm (SIDIS)}(z,\mu_B)&= \delta_{q'q} [\delta(1-z)+\frac{\alpha_s}{\pi}(\frac{C_F}{2}(1-z)  -2C_F\delta(1-z) + P_{q\gets q}(z)\, \ln z) ]\; ,  \label{eq:cd_css}
\\
\hat C_{g\gets q}^{\rm (SIDIS)}(z,\mu_B)&= \frac{\alpha_s}{\pi} \left( \frac{C_F}{2} z\; +  P_{g\gets q}(z)\, \ln z \right),
\label{eq:cd1_css}
\end{align}
where $\alpha_s$ represents the strong coupling, and the expansion at the next-to-leading order can be written as follows
\begin{align}
\alpha_{s}\left(Q^{2}\right)=\frac{12 \pi}{\left(33-2 n_{f}\right) \ln \left(Q^{2} / \Lambda_{Q C D}^{2}\right)}\left\{1-\frac{6\left(153-19 n_{f}\right)}{\left(33-2 n_{f}\right)^{2}} \frac{\ln \ln \left(Q^{2} / \Lambda_{Q C D}^{2}\right)}{\ln \left(Q^{2} / \Lambda_{Q C D}^{2}\right)}\right\}.
\label{eq:alpha}
\end{align}
In Eq.~(\ref{eq:alpha}), $Q^{2}$ is the running energy scale and $n_{f}=5$, $\Lambda_{Q C D}=0.225$ GeV.
The splitting functions $P_{q\gets q}$ and $P_{g\gets q}$ in Eq.~(\ref{eq:cd_css}) and Eq.~(\ref{eq:cd1_css})  have the general form
\begin{align}
P_{q\gets q}(z) &= C_F \left[ \frac{1+z^2}{(1-z)_+} + \frac{3}{2} \delta(1-z) \right],
\label{P_qq}
\\
P_{g\gets q}(z) &= C_F \frac{1+(1-z)^2}{z},
\label{P_gq}
\end{align}
where $C_F=4/3$, $T_R=1/2$, and the subscript symbol ``$+$'' denotes the following prescription
\begin{align}
\int_0^1 dz {f(z)\over (1-z)_+} =\int_0^1 dz {f(z)-f(1)\over (1-z)}
\end{align}
Combining the Bessel function part, the PDF part $\tilde{f}_1^{q/p}(x,b)$ and the FF part $\tilde{D}_1^q(z,b)$, we can expand the denominator of Sivers asymmetry as
\begin{align}
F_{UU}(Q;P_{hT})
&=\frac{x}{z^{2}} \sum_{q} e_{q}^{2}\int_{0}^{\infty}\frac{bd b }{(2 \pi)} J_{0}\left(\frac{P_{h T}b}{z}\right) e^{-{S}_{\rm pert}(Q;b_*)-S_{\rm NP}^{\rm SIDIS}(Q;b)}\nonumber\\
&~~\left( \sum_i C\int_{x}^{1} \frac{d \xi}{\xi} C_{q \leftarrow i}^{\rm (SIDIS)}\left(\frac{x}{\xi},  \mu_{B}\right) f_{1}^{i/p}(\xi, \mu_{B}) \right) \times\left(\sum_j \int_{z}^{1} \frac{d\xi}{\xi} \hat{C}_{j \leftarrow q}^{\rm (SIDIS)}\left(\frac{z}{\xi}, \mu_{B}\right) D_1^{h / j}\left(\xi, \mu_{B}\right)\right),
\label{fuu_res}
\end{align}
where the nonperturbative Sudakov-like form factor $S_{\mathrm{NP}}^{\operatorname{SIDIS}}(Q ; b)$ receives the contribution from the unpolarized TMD PDF and FF:
\begin{align}
S_{\mathrm{NP}}^{\operatorname{SIDIS}}(Q ; b) &=S_{\mathrm{NP}}^{\mathrm{f}_{1}}(Q ; b)+S_{\mathrm{NP}}^{\mathrm{D}_{1}}(Q ; b) \nonumber\\
&=g_{2} \ln \left(\frac{Q}{Q_{0}}\right) b^{2}+g_{1}^{\mathrm{f}_{1}} b^{2}+g_{1}^{\mathrm{D}_{1}} b^{2}.
\label{snpsidis}
\end{align}
Here, $g_{1}^{\mathrm{f}_{1}}$ and $g_{1}^{\mathrm{D}_{1}}$ are obtained from Eq.~(\ref{eq:g1}), so $g_{1}^{\mathrm{f}_{1}}=g_1^{\rm pdf} = \frac{\langle p_\perp^2\rangle_{Q_0}}{4}$,  $g_{1}^{\mathrm{D}_{1}}=g_1^{\rm ff} = \frac{\langle k_\perp^2\rangle_{Q_0}}{4z_h^2}$.
	
\subsection{The solution of the transverse spin-dependent structure function}

Similar to the previous subsection, we further obtain the expression for the transverse spin-dependent structure function $F_{UT}^{\sin(\phi_h-\phi_s)}$.
Eq.~(\ref{futt_expanded}) can be separated to three parts to solve it, namely, the weighted Bessel function, the Qiu-Sterman~(QS) function $T_{q, F}\left(x^{\prime}, x^{\prime \prime}\right)$ and the fragmentation function $\tilde{D}_1^{h/q}(z,b)$.
The solution for the weighted Bessel function is similar to Eq.~(\ref{eq:fuucoe})
\begin{align}
-\frac{x}{2 z^{2}} \sum_{q} e_{q}^{2}\int \frac{d^{2} b}{(2 \pi)^{2}} i \hat{h}_\alpha b^{\alpha} e^{i \bm{P_{h T}} \cdot \boldsymbol{b} / z }
&=\frac{x}{2 z^{2}} \sum_{q} e_{q}^{2}\int_{0}^{\infty}\frac{b^{2}db}{2 \pi} J_{1}\left(\frac{P_{h T}b}{z}\right).
\end{align}
The QS function part $T_{q, F}\left(x^{\prime}, x^{\prime \prime}\right)$ in the $b$ space has a similar solution form after solving the TMD evolution equations
\begin{align}
T_{q,F}\left(x, x;Q \right)=\mathcal{T} \times e^{-S} \times T_{q,F}\left(x, x;\mu_{B} \right).
\end{align}
In the small $b$ region, the QS function can be calculated from the convolution of the hard scattering coefficients and the collinear counterpart utilizing the perturbative QCD
\begin{align}
T_{q,F}\left(x, x;\mu_{B} \right)=  \Delta C_{q \leftarrow i}^{T} \otimes T_{i, F}\left(x, x ; \mu_{B}\right),
\end{align}
where $\otimes$ denotes convolution and can be expended as
\begin{align}
\Delta C_{q \leftarrow i}^{T} \otimes T_{i, F}\left(x, x ; \mu_{B}\right)= \sum_{i}\int_{x}^{1} \frac{d\xi}{\xi} \Delta C_{q \leftarrow i}^{T}\left(\frac{x}{\xi}, \mu_{B}\right) T_{i, F}\left(\xi, \xi ; \mu_{B}\right),
\end{align}
thus the QS function $T_{q, F}\left(x^{\prime}, x^{\prime \prime}\right)$ and the fragmentation function $\tilde{D}_1^{h/q}(z,b)$ can be expanded as
\begin{align}
T_{q,F}\left(x, x;Q \right)=e^{-\frac{1}{2}S_{\mathrm{Pert}}(Q;b_\ast)
-S^{f_1}_{\mathrm{NP}}(Q;b)}\mathcal{T}(\alpha_s(Q))
\sum_i\int_{x}^{1} \frac{d\xi}{\xi} \Delta C_{q \leftarrow i}^{T}\left(\frac{x}{\xi}, \mu_{B}\right) T_{i, F}\left(\xi, \xi; \mu_{B}\right),\label{Tqf}\\
\tilde{D}_1^{h/q}(z,b;Q)=e^{-\frac{1}{2}S_{\mathrm{Pert}}(Q;b_\ast)
-S^{D_1}_{\mathrm{NP}}(Q;b)}\mathcal{D}(\alpha_s(Q))
\sum_j\int_{z}^{1} \frac{d\xi}{\xi} \hat{C}_{j \leftarrow q}\left(\frac{z}{\xi}, \mu_{B}\right) D_1^{h / j}\left(\xi, \mu_{B}\right),
\end{align}
where the Sudakov-like form factor is expressed in Eq.~(\ref{eq:Spert}), Eq.~(\ref{eq:snppdf}) and Eq.~(\ref{eq:snpff}).
And the hard scattering coefficient $\mathcal{T}(\alpha_s(Q))$ and $\mathcal{D}(\alpha_s(Q))$ equal to 1.
The $C$ coefficient o $\tilde{D}_1^q(z,b)$ is expressed in Eqs.~(\ref{eq:cd_css}), (\ref{eq:cd1_css}), (\ref{P_qq}) and (\ref{P_gq}).
The expression of $\Delta C$ in Eq.~(\ref{Tqf}) is~\cite{Sun:2013hua}
\begin{align}
\Delta C_{q q}^{T(D I S)}(x,\mu_{B})=\delta(1-x)+\frac{\alpha_{s}(\mu_{B})}{\pi}\left(-\frac{1}{4 N_{c}}(1-x)-2 C_{F} \delta(1-x)\right),
\end{align}
where $N_{c}=3$. 
Combining all the ingredients together, the numerator part of the Sivers asymmetry can be rewritten as
\begin{align}
F_{U T, T}^{\sin \left(\phi_{h}-\phi_{S}\right)}=& -\frac{x}{2z^{2}} \sum_{q} e_{q}^{2} \int \frac{d^{2} b}{(2 \pi)^{2}} e^{i \boldsymbol{P}_{h T} / z \cdot \boldsymbol{b}} \hat{h}_\alpha\cdot b^{\alpha} T_{q,F}\left(x, x\right) \tilde{D}_1^{h/q}(z,b)\nonumber\\
=& \frac{x}{2z^{2}} \sum_{q} e_{q}^{2} \int_{0}^{\infty}\frac{b^{2}db}{2 \pi} J_{1}\left(\frac{P_{h T}b}{z}\right) e^{-S_{\text {pert }}\left(Q ; b_{*}\right)-S_{\mathrm{NP\ Sivers}}^{\text {SIDIS }}(Q ; b)} \nonumber\\
&\left(\sum_{i} \int_{x}^{1} \frac{d\xi}{\xi} \Delta C_{q \leftarrow i}^{T(D I S)}\left(\frac{x}{\xi}, \mu_{B}\right) T_{i, F}\left(\xi, \xi ; \mu_{B}\right)\right) \times\left(\sum_{j} \int_{z}^{1} \frac{d\xi}{\xi} \hat{C}_{j \leftarrow q}^{(SIDIS)}\left(\frac{z}{\xi}, \mu_{B}\right) D_1^{h / j}\left(\xi, \mu_{B}\right)\right),
\label{futtresult}
\end{align}
where the nonperturbative Sudakov-like form factor $S_{\mathrm{NP\ Sivers}}^{\operatorname{SIDIS}}(Q ; b)$ is the combination of and the one for the Sivers function the one for the unpolarized TMD FF
\begin{align}
S_{\mathrm{NP\ Sivers}}^{\operatorname{SIDIS}}(Q ; b)&=S_{\mathrm{NP}}^{Sivers}(Q ; b)+S_{\mathrm{NP}}^{\mathrm{D}_{1}}(Q ; b)  \nonumber\\
&=g_{2} \ln \left(\frac{Q}{Q_{0}}\right) b^{2}+g_{1}^{Sivers} b^{2}+g_{1}^{\mathrm{D}_{1}} b^{2},
\end{align}
where $g_{1}^{Sivers}=\frac{\langle k_{s\perp}^2\rangle}{4}$, $g_{1}^{\mathrm{D}_{1}}=\frac{\langle k_\perp^2\rangle_{Q_0}}{4z_h^2}$.

\section{NUMERICAL ESTIMATE}
\label{sec:number}
In this section, we present the numerical estimate for the Sivers asymmetry in charged Kaon and $\Lambda$ hyperon produced SIDIS process at the kinematical configurations of EIC and EicC.

In order to obtain the results for the Sivers asymmetry, we shall
have the collinear distribution functions and fragmentation functions as input. For the
unpolarized  proton collinear distribution function $f_{1}(x,\mu_{B})$, we apply the parametrization from MSTW2008~\cite{Martin:2009iq}.
For the collinear unpolarized fragmentation function $D_{1}(z)$ of charged Kaon, the DSS parameterization results at NLO accuracy is used~\cite{deFlorian:2007aj}.
For the collinear unpolarized fragmentation function $D_{1}^{\Lambda}(z)$ of $\Lambda$ hyperon, we adopt the model results from the diquark spectator model~\cite{Yang:2017cwi}
\begin{align}
D_{1}^{\Lambda}(z)=&\frac{g_{s}^{2}}{4(2 \pi)^{2}} \frac{e^{-\frac{2m_{q}^{2}}{\Lambda^{2}}}}{z^{4} L^{2}}\left\{z(1-z)\left(\left(m_{q}+M_{\Lambda}\right)^{2}-m_{D}^{2}\right)\right.\times \exp \left(\frac{-2 z L^{2}}{(1-z) \Lambda^{2}}\right)\nonumber\\
&+\left((1-z) \Lambda^{2}-2\left(\left(m_{q}+M_{\Lambda}\right)^{2}-m_{D}^{2}\right)\right)\left.\times \frac{z^{2} L^{2}}{\Lambda^{2}} \Gamma\left(0, \frac{2 z L^{2}}{(1-z) \Lambda^{2}}\right)\right\}.
\label{D1Lambda}
\end{align}
The values of the free parameters in Eq.~(\ref{D1Lambda}) are taken from Ref.~\cite{Yang:2017cwi}.
Since the model result is obtained at the initial energy of 0.23 GeV$^{2}$, to make it applicable to a more general energy range, we use the QCDNUM evolution package~\cite{Botje:2010ay} to evolve the unpolarized fragmentation function $D_{1}(z)$ from the initial energy of 0.23 GeV$^{2}$ to another energy scale.

In our calculation we adopt the Qiu-Sterman~(QS) function $T_{q, F}\left(x, x ; \mu_{B}\right)$ from the parameterization in Ref.~\cite{Echevarria:2014xaa}
\begin{align}
T_{q, F}(x, x, \mu_{B})=N_{q} \frac{\left(\alpha_{q}+\beta_{q}\right)^{\left(\alpha_{q}+\beta_{q}\right)}}{\alpha_{q}^{\alpha_{q}} \beta_{q}^{\beta^{q}}} x^{\alpha_{q}}(1-x)^{\beta_{q}} f_{1}(x, \mu_{B}).
\label{QS}
\end{align}
The parameters obtained by fitting to the data in SIDIS processes from HERMES, COMPASS and JLab ~\cite{HERMES:2009lmz,COMPASS:2008isr,COMPASS:2012dmt,JeffersonLabHallA:2011ayy}, are listed in Table.~\ref{tab:qs}.
\begin{table}[h]
\centering
\normalsize
\setlength{\tabcolsep}{4mm}{
\begin{tabular}{|c|c|c|c|c|c|}
\hline
$\alpha_{u}$&  $\alpha_{d}$&  $\alpha_{sea}$&  $\beta$& $N_{u}$ &$N_{d}$ \\
\hline
1.051&  1.552&  0.851&  4.857&  0.106& -0.163 \\
\hline
$N_{\bar{u}}$&  $N_{\bar{d}}$&  $N_{s}$&  $N_{\bar{s}}$&  $\langle k_{s\perp}^2\rangle$&  \\
\hline
-0.012&  -0.105&  0.103&  -1.000&  0.282 $\mathrm{GeV}^{2}$&  \\
\hline
\end{tabular}}
\caption{Values of the parameters in the parameterization of the Qiu-Sterman function in Ref.~\cite{Echevarria:2014xaa}}\label{tab:qs}
\end{table}

 We should notice that in Ref.~\cite{Echevarria:2014xaa}, the QS function is assumed to be  proportional to the collinear unpolarized distribution function $f_{1}(x,\mu_{B})$ in which the DGLAP evolution effect is not included.
 To investigate the impact of the QS function DGLAP evolution effect on the Sivers asymmetry, we take $Q_{0}^{2}=2.4 ~\mathrm{GeV}^{2}$ as the initial energy and evolve the QS function in Eq.~(\ref{QS}) into another energy.
 We adopt two different approaches to evolve the QS function: one is to assume the QS function follows the same evolution effect as that for unpolarized distribution function, the other one is to change the evolution kernel in the QCDNUM evolution package to include the QS evolution kernel by considering the homogenous terms [the terms containing $T_{q, F}(x, x, \mu_{B})$] in the evolution kernel as an approximation~\cite{Kang:2008ey}:
\begin{align}
P_{q q}^{\mathrm{Sivers}} \approx \frac{4}{3}\left(\frac{1+z^2}{(1-z)_{+}}+\frac{3}{2} \delta(1-z)\right)-\frac{3}{2} \frac{1+z^2}{1-z}-3 \delta(1-z).
\label{siversevo}
\end{align}

In Eq.~(\ref{snp}) and Eq.~(\ref{eq:g1}), the free parameter $g_{1}$ and the universal parameter $g_{2}$ contain information about the evolution of TMDs and are the key parameters that determine the evolution of TMDs from one initial energy $\mu$ to another $Q$.
Here we adopt the results given in Ref.~\cite{Echevarria:2014xaa} for the mean transverse momentum squared $$\left\langle p_{\perp}^{2}\right\rangle=0.38 ~\mathrm{GeV}^{2},\left\langle k_{\perp}^{2}\right\rangle=0.19 ~\mathrm{GeV}^{2}.$$
For the universal parameter $g_{2}$ in the nonperturbative Sudakov-like form factor, the specific value $g_{2}=0.16$ is also given in Ref.~\cite{Echevarria:2014xaa}.

The kinematical region that is available at EIC is chosen as follows~\cite{Accardi:2012qut}
\begin{align}
0.001<x<0.4, \quad 0.07<y&<0.9, \quad 0.2<z<0.8 \text {, }\nonumber\\
1 \mathrm{GeV}^{2}<Q^{2}, \quad W>5 \mathrm{GeV}, \quad \sqrt{s}&=100 \mathrm{GeV}, \quad P_{h T}<0.5 \mathrm{GeV} .
\end{align}
As for the EicC, the following kinematical cuts are adopted
\begin{align}
0.005<x<0.5, \quad 0.07<&y<0.9, \quad 0.2<z<0.7, \nonumber\\
1 \mathrm{GeV}^{2}<Q^{2}<200 \mathrm{GeV}^{2}, \quad W>2 \mathrm{GeV}&, \quad \sqrt{s}=16.7 \mathrm{GeV}, \quad P_{h T}<0.5 \mathrm{GeV},
\end{align}
where $W^{2}=(P+q)^{2} \approx \frac{1-x}{x} Q^{2}$ is invariant mass of the virtual photon-nucleon system.
Since TMD factorization is proved to be valid to describe the physical observables in the region $P_{h T} \ll Q$, $P_{h T}<0.5 ~\mathrm{GeV}$ is chosen to guarantee the validity of TMD factorization.
Combining Eq.~(\ref{eq:asymmetry}), Eq.~(\ref{fuu_res}) and Eq.~(\ref{futtresult}) and the kinematical regions of EIC and EicC, we can calculate the single-spin dependent Sivers asymmetry of charged Kaon and $\Lambda$ hyperon produced SIDIS process within the EIC and EicC kinematical region.

\begin{figure}[h]
	\centering
	\includegraphics[width=0.3\columnwidth]{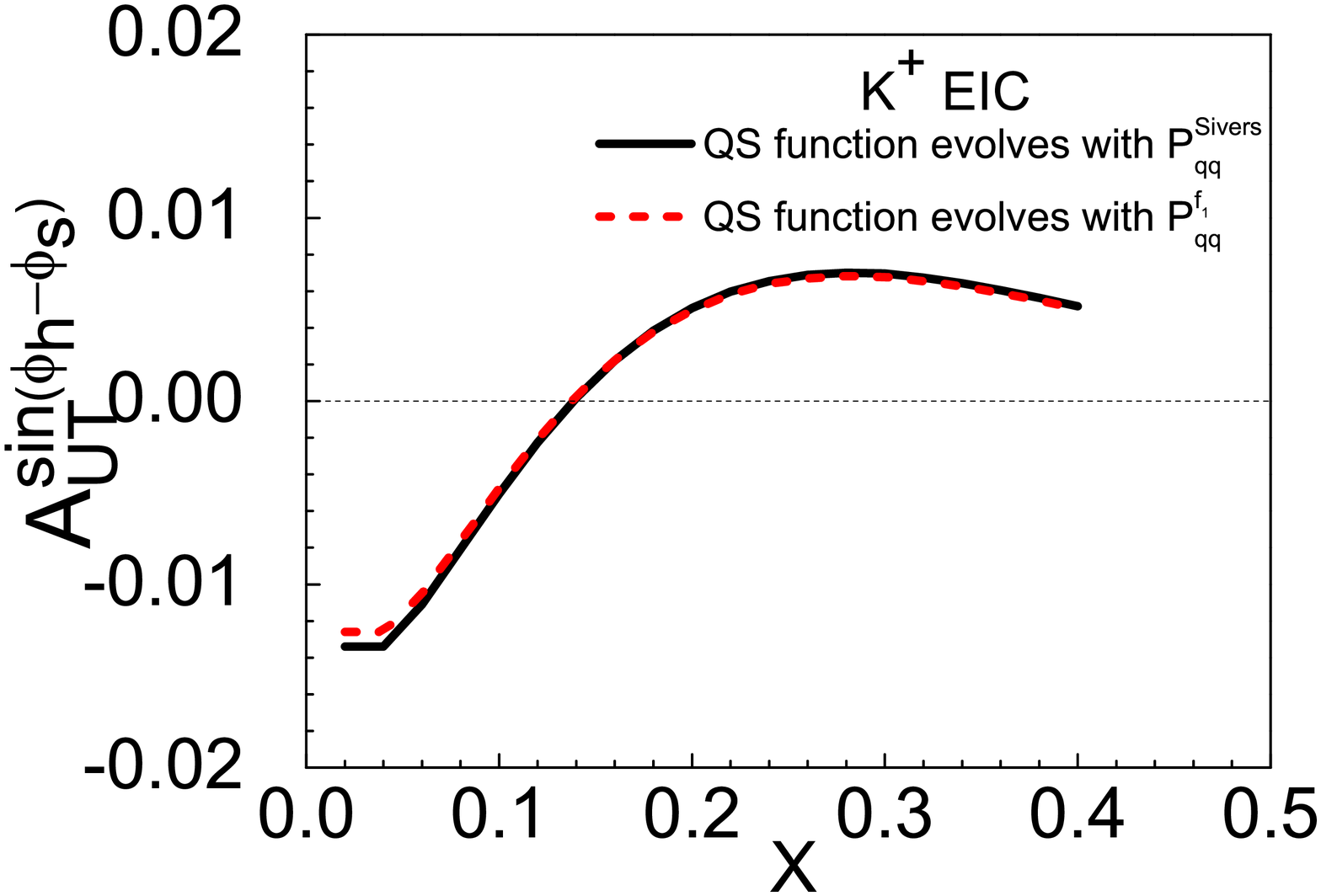}
	\includegraphics[width=0.3\columnwidth]{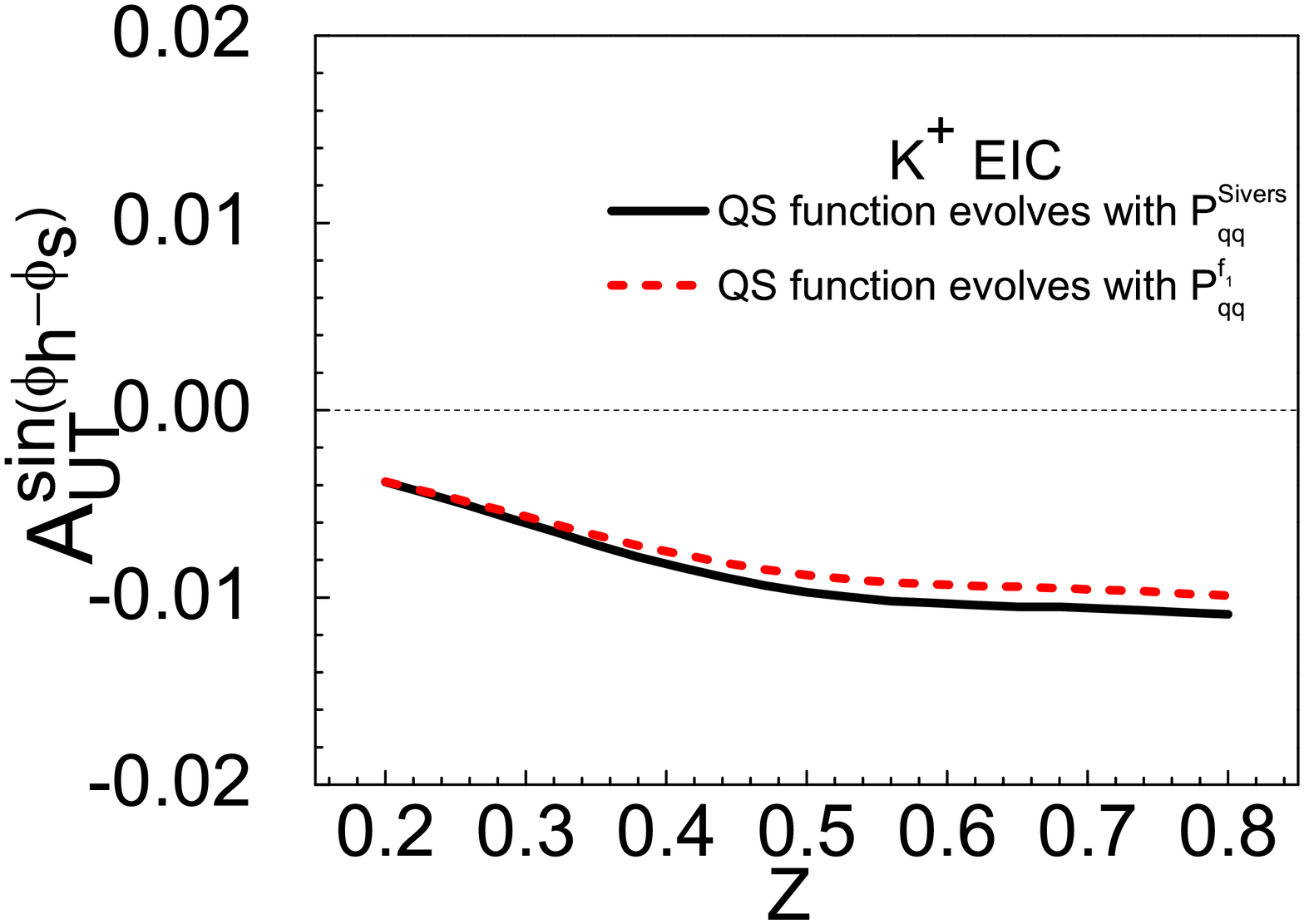}
	\includegraphics[width=0.3\columnwidth]{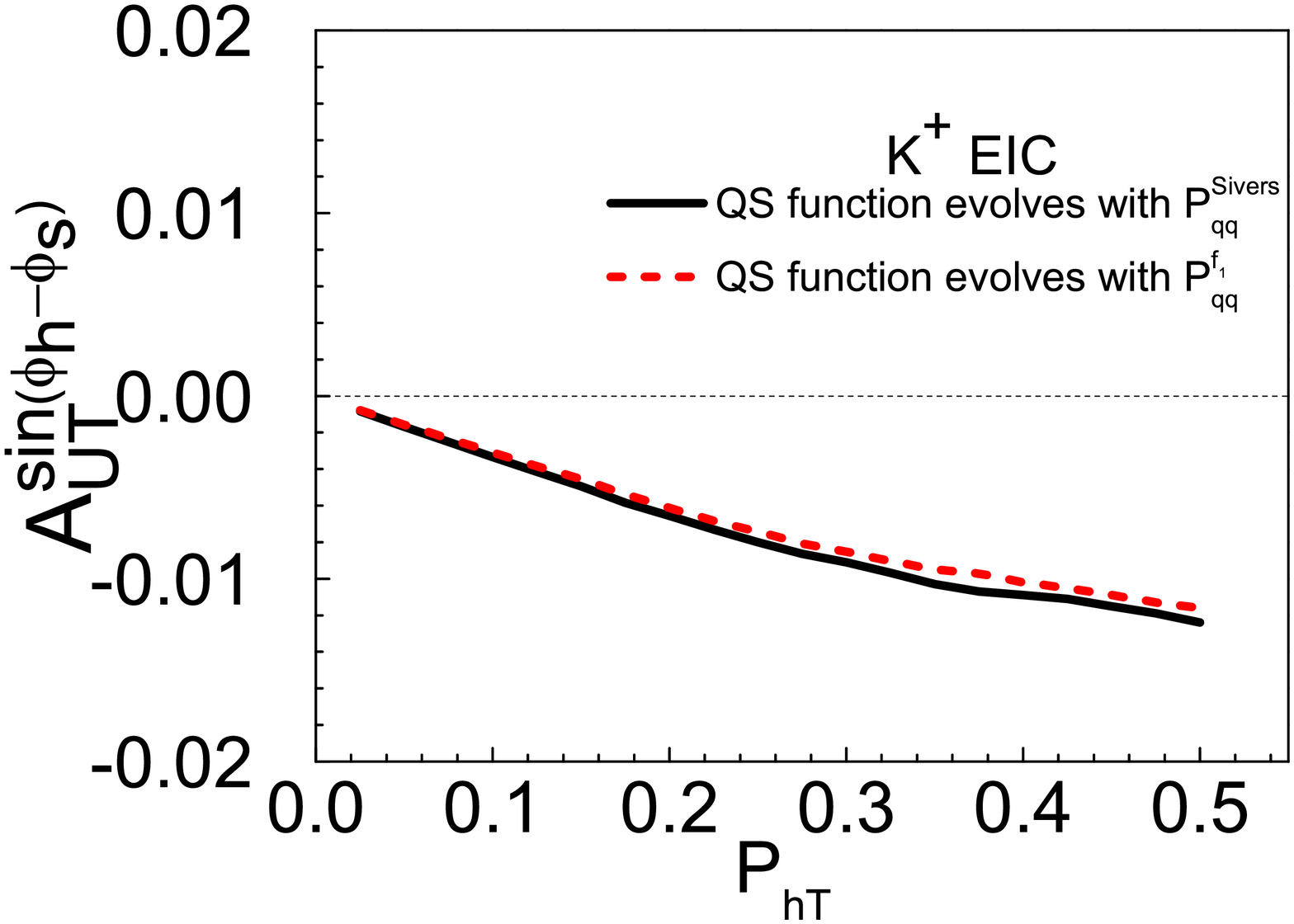}
	\includegraphics[width=0.3\columnwidth]{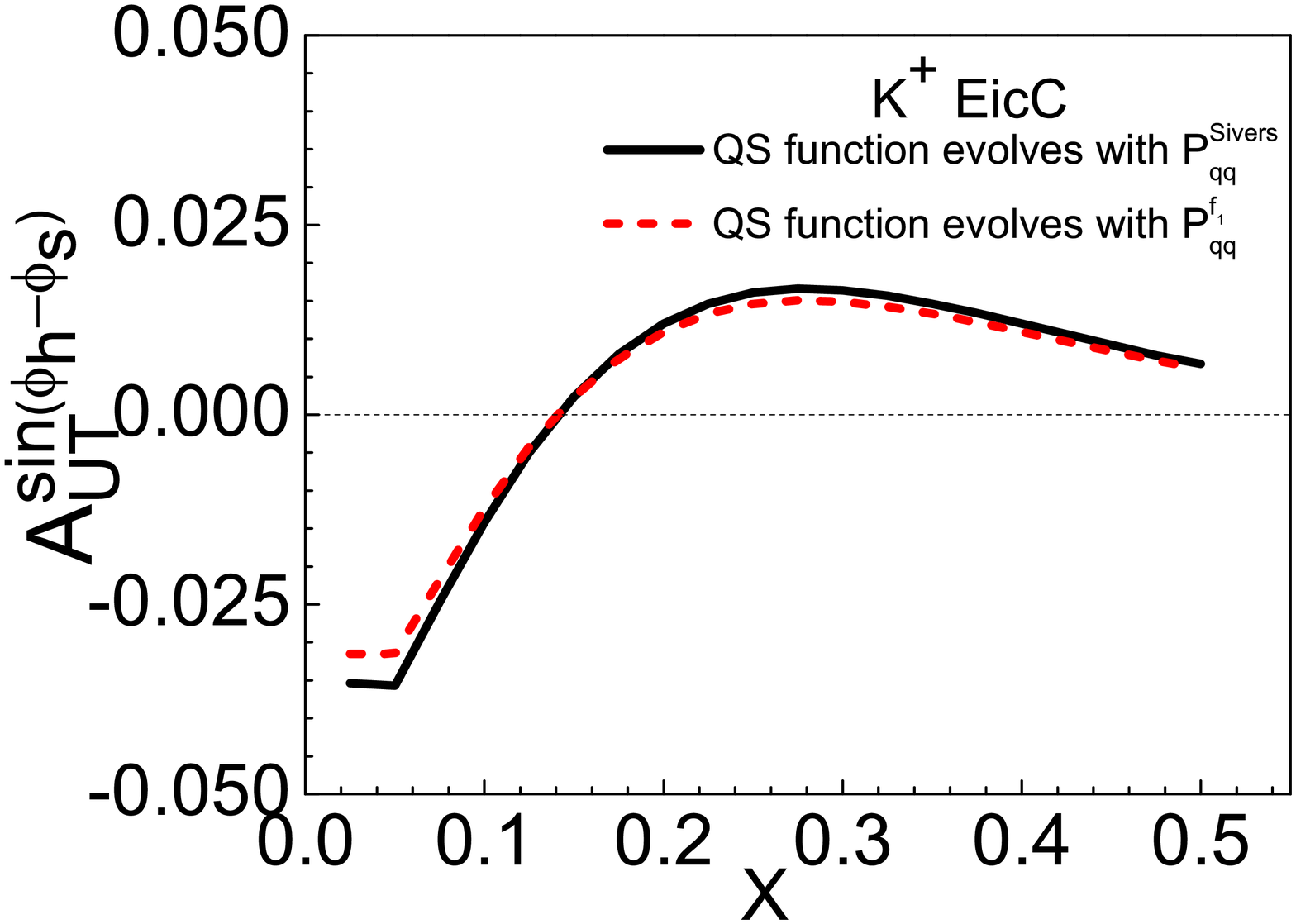}
	\includegraphics[width=0.3\columnwidth]{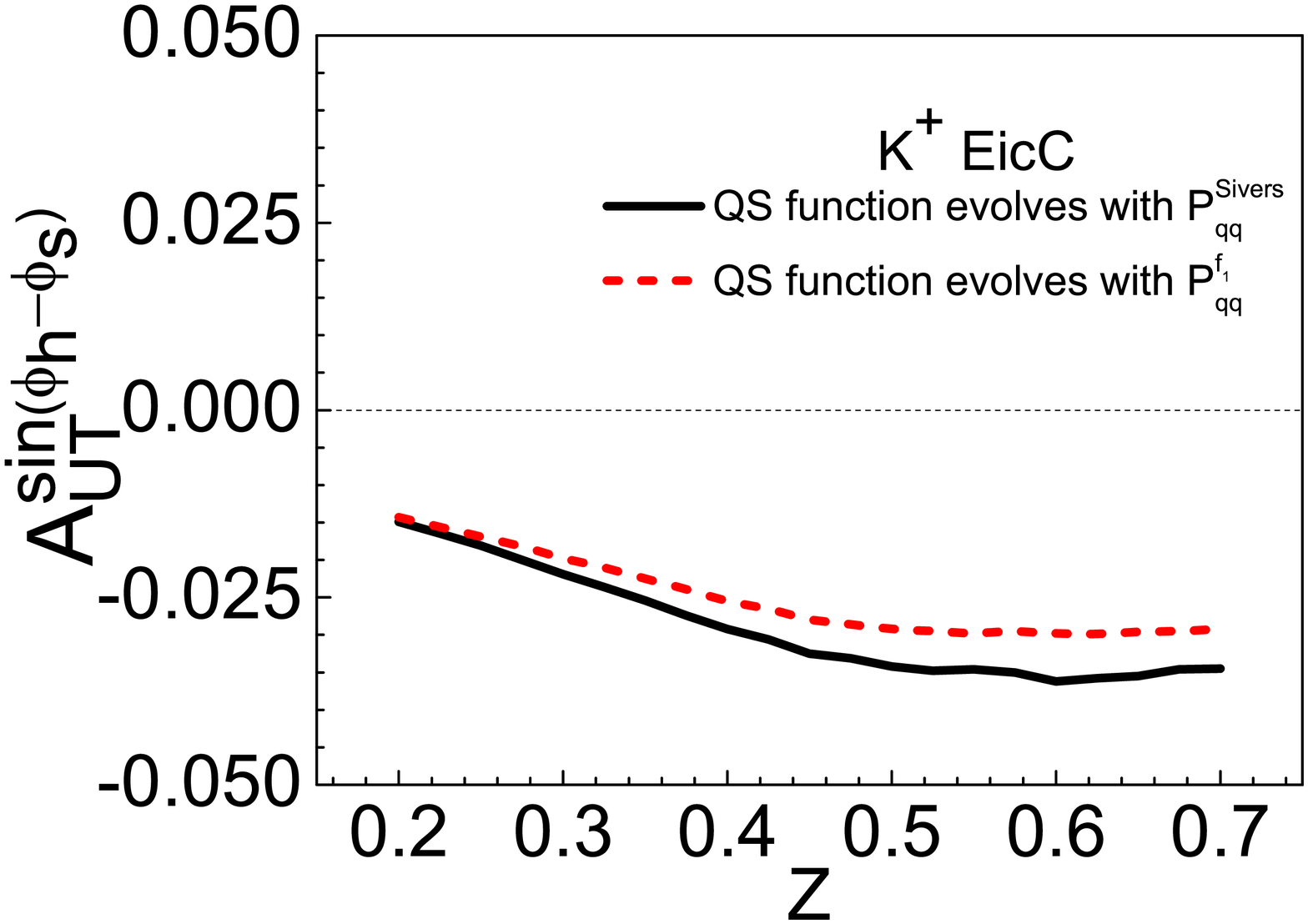}
	\includegraphics[width=0.3\columnwidth]{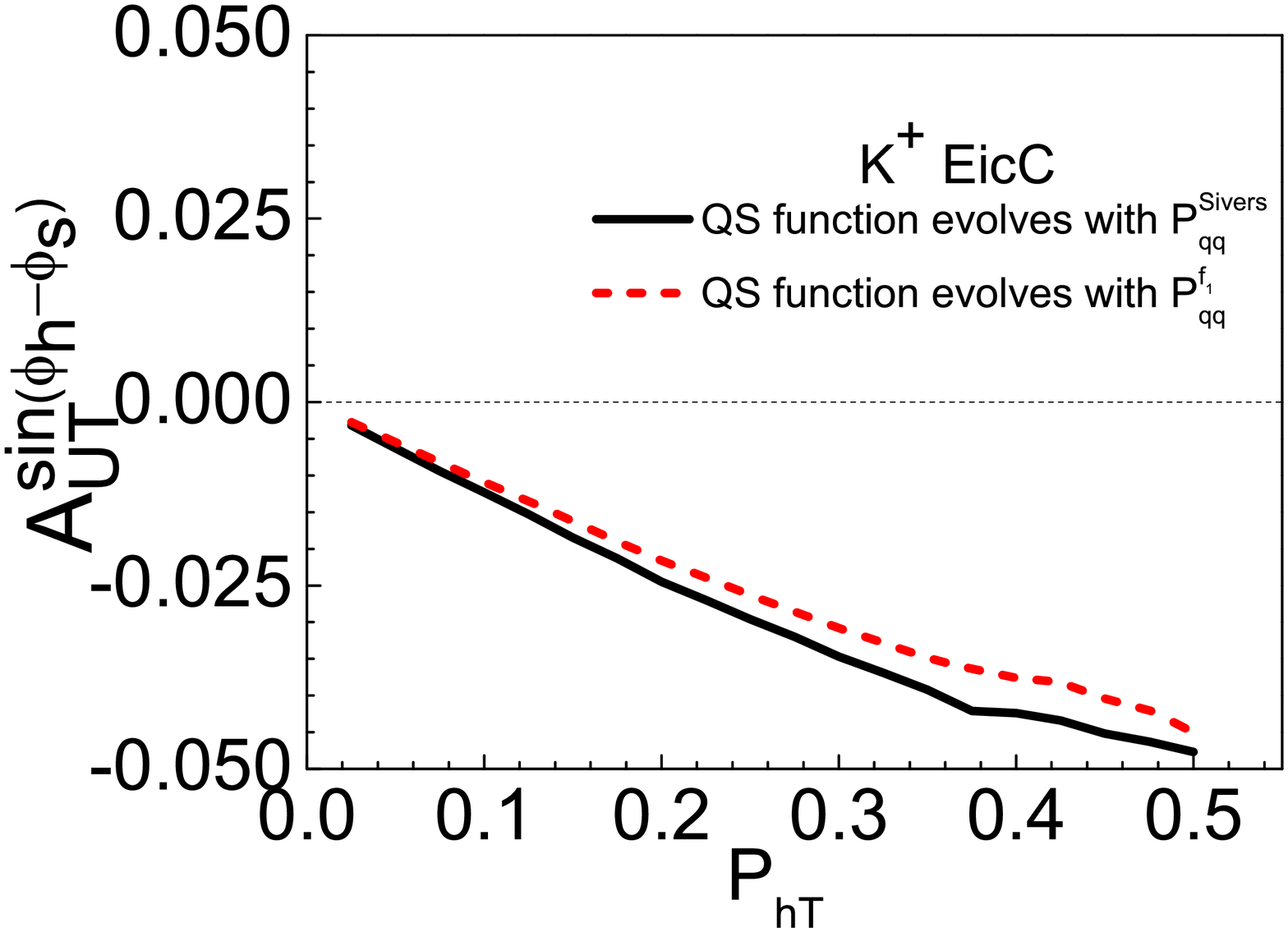}
	\includegraphics[width=0.3\columnwidth]{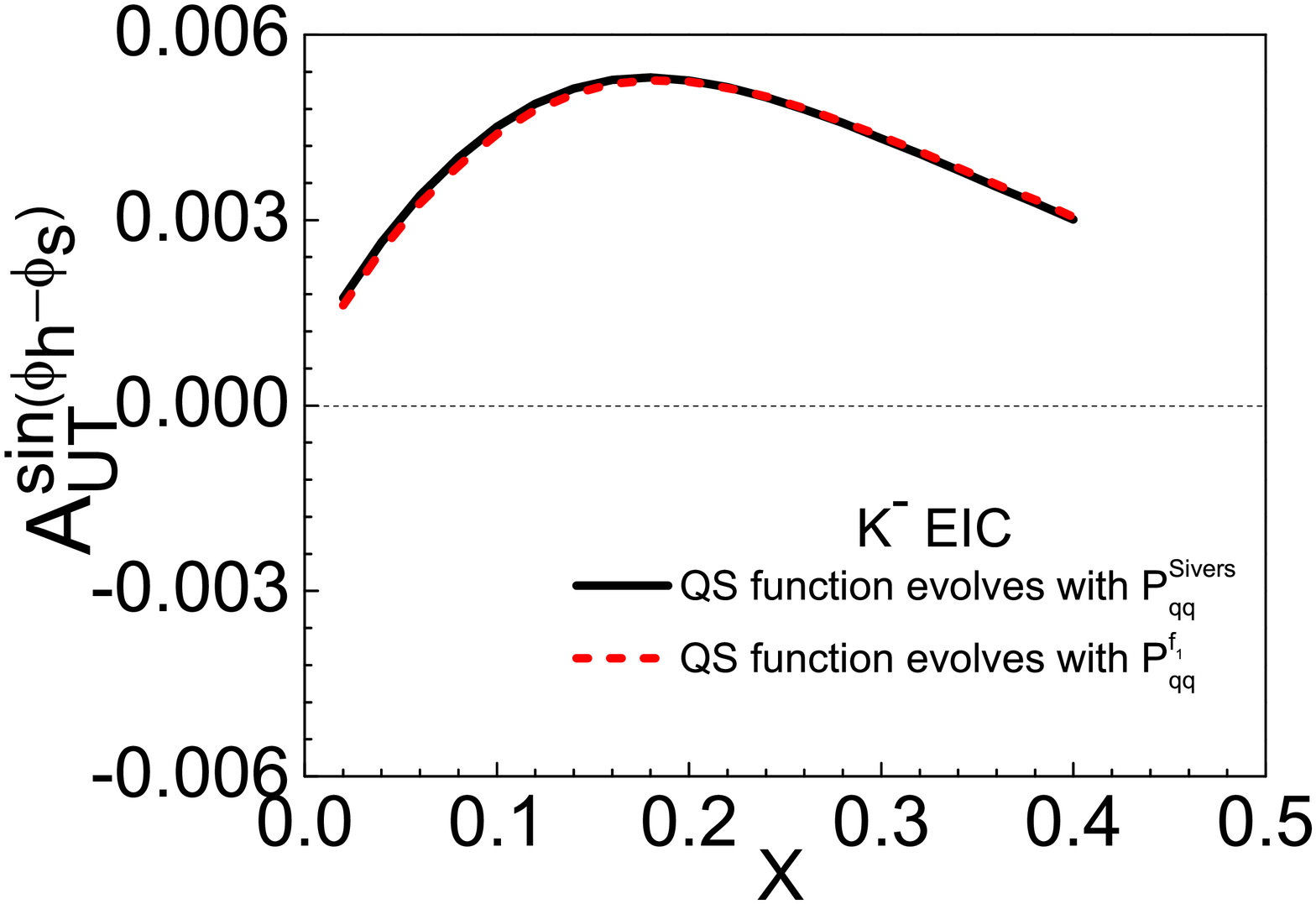}
	\includegraphics[width=0.3\columnwidth]{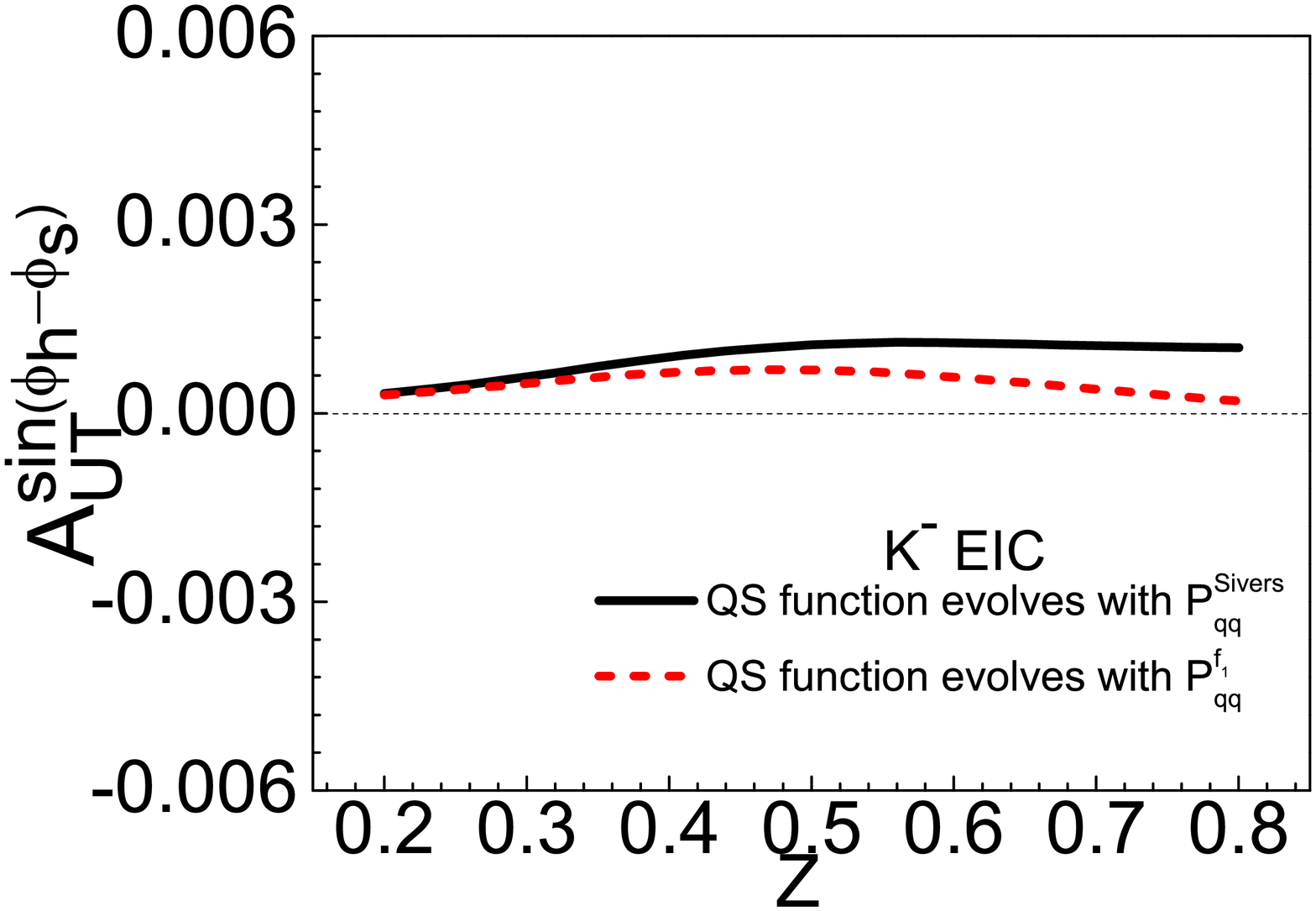}
	\includegraphics[width=0.3\columnwidth]{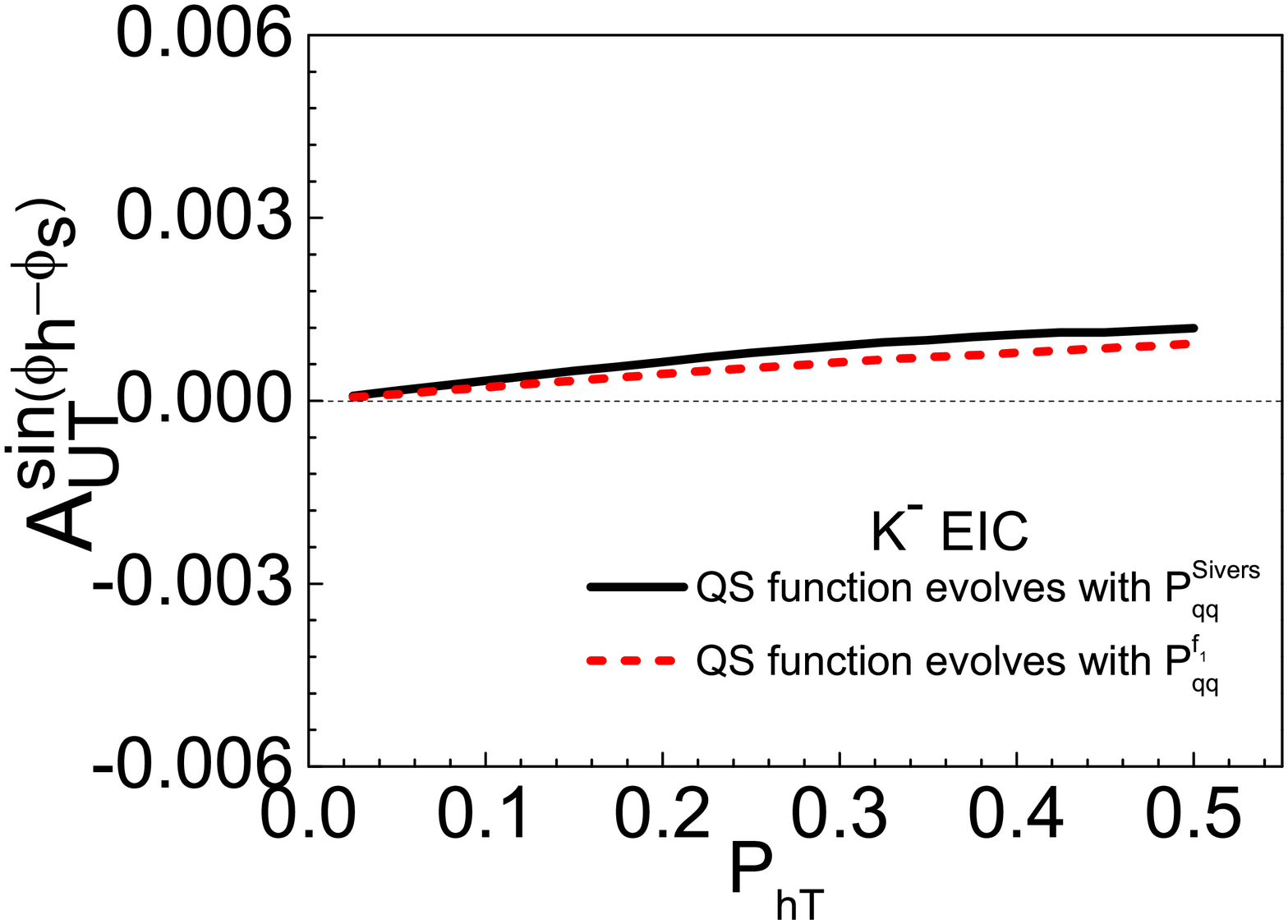}
	\includegraphics[width=0.3\columnwidth]{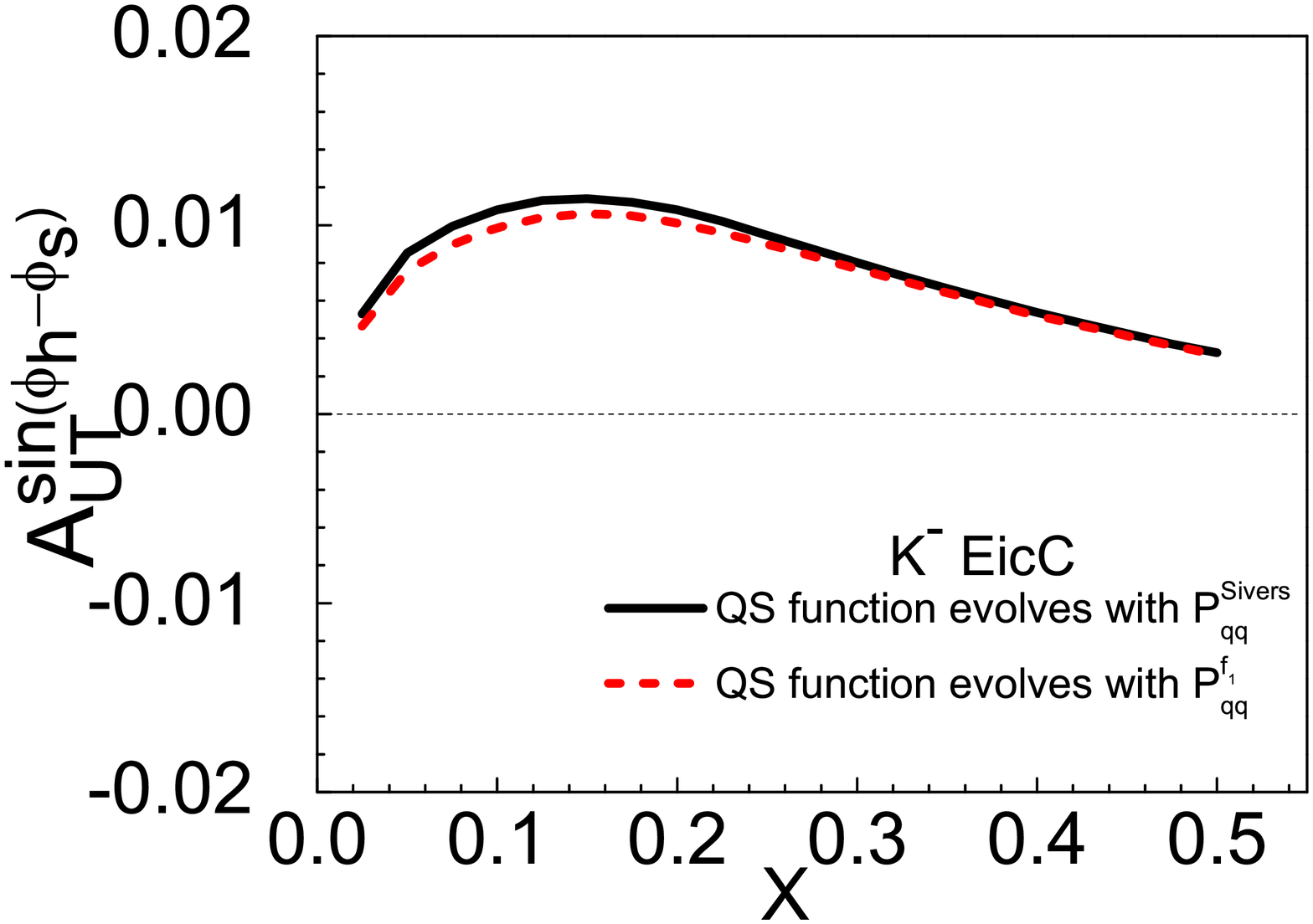}
	\includegraphics[width=0.3\columnwidth]{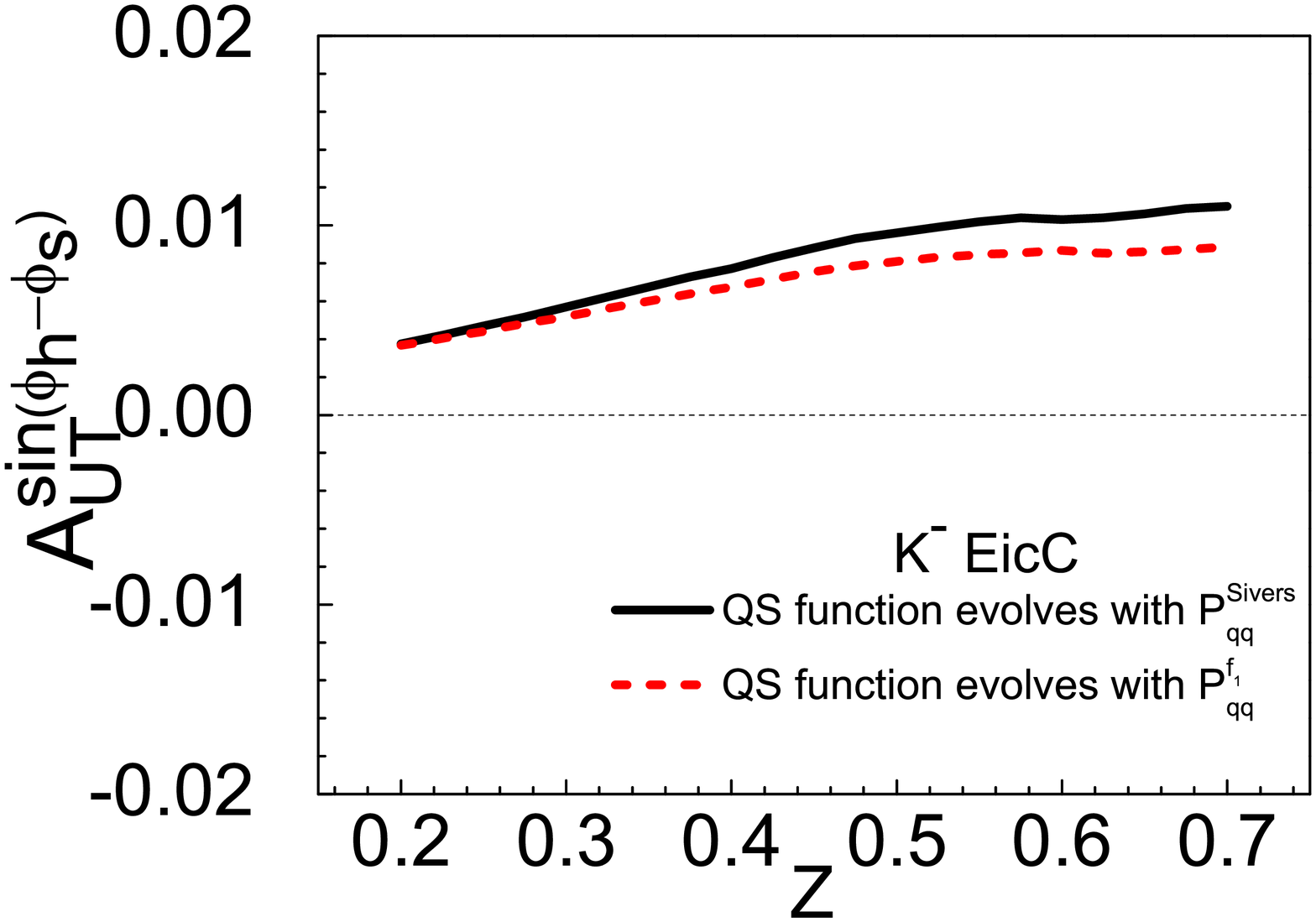}
	\includegraphics[width=0.3\columnwidth]{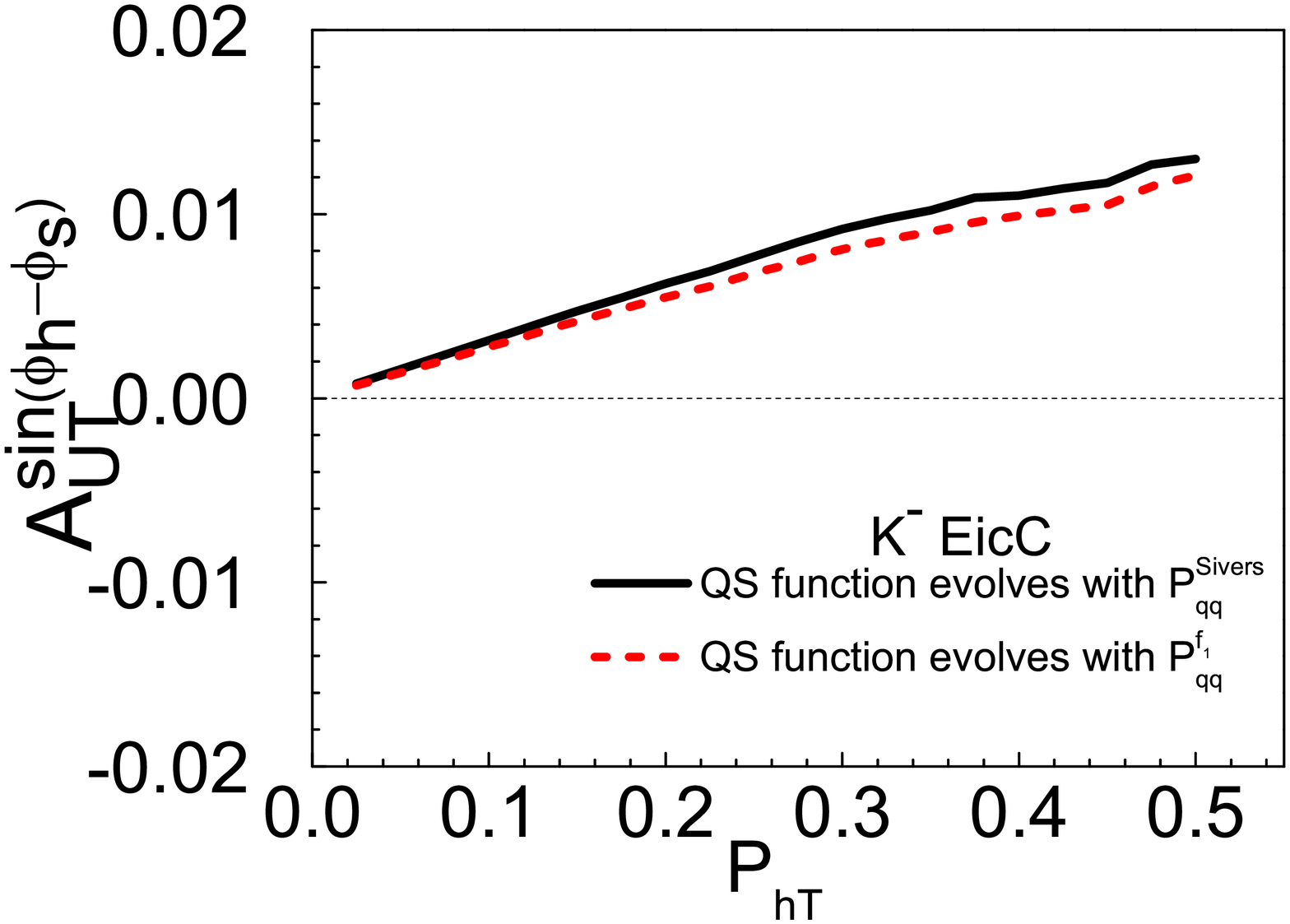}
	\includegraphics[width=0.3\columnwidth]{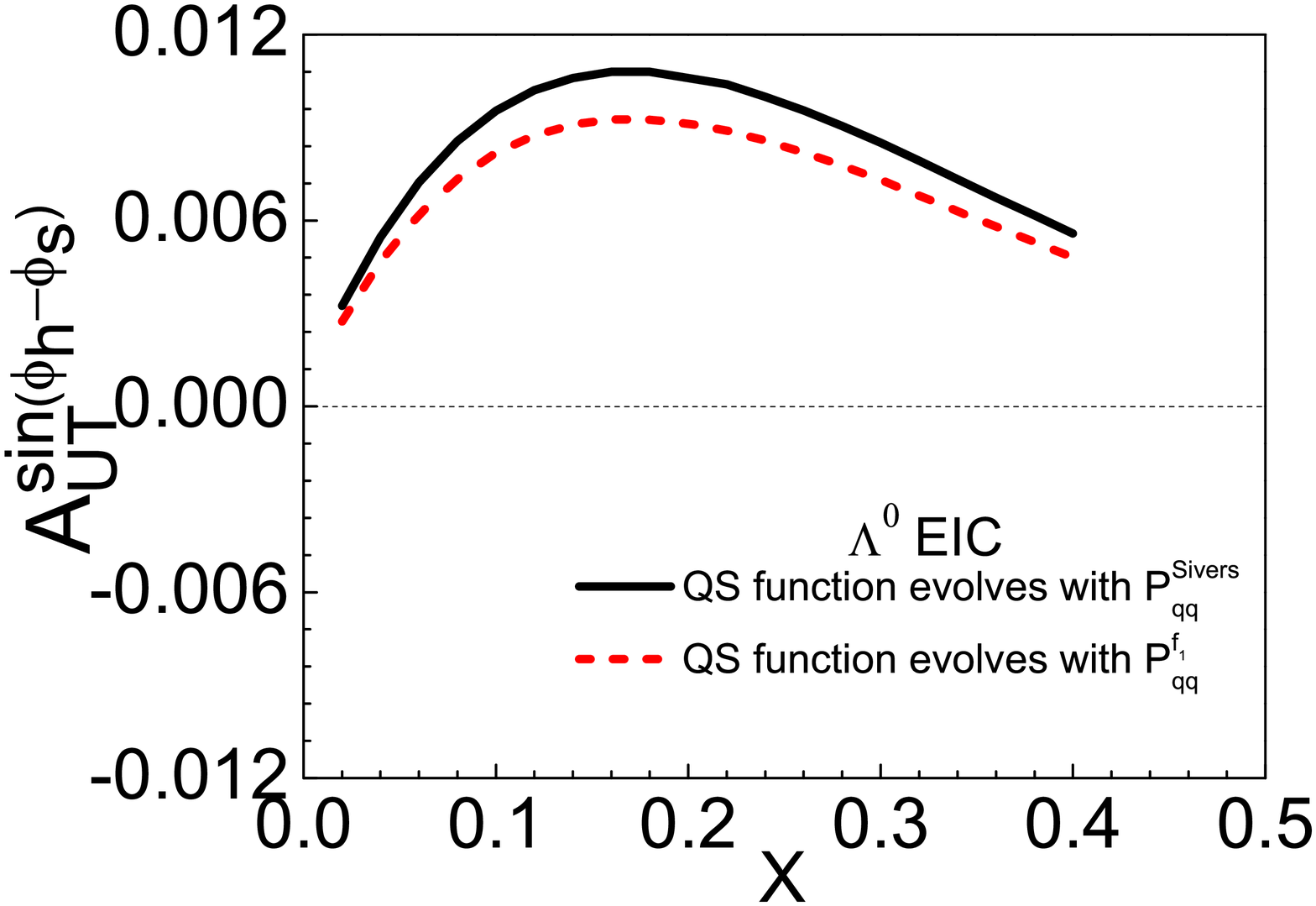}
	\includegraphics[width=0.3\columnwidth]{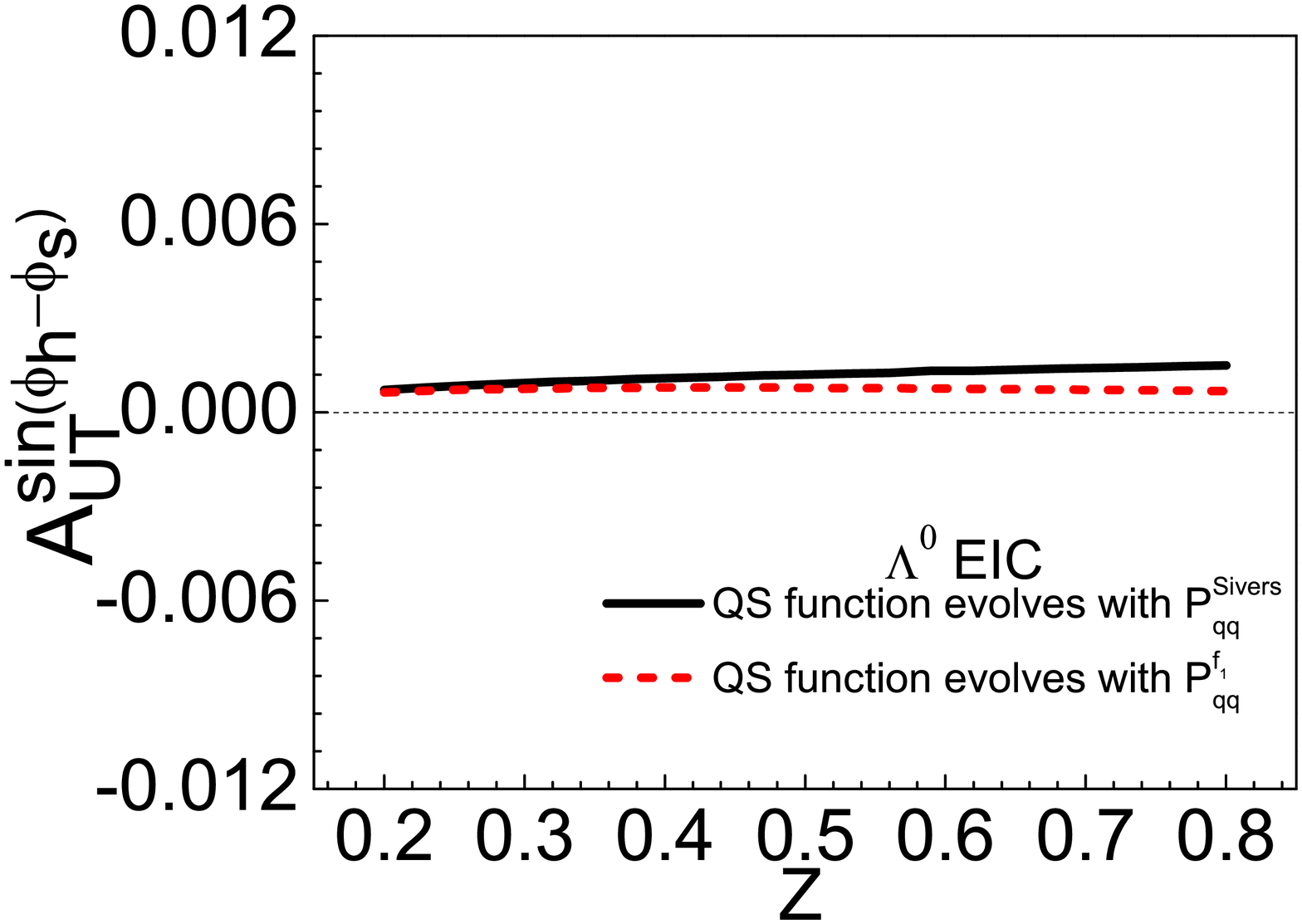}
	\includegraphics[width=0.3\columnwidth]{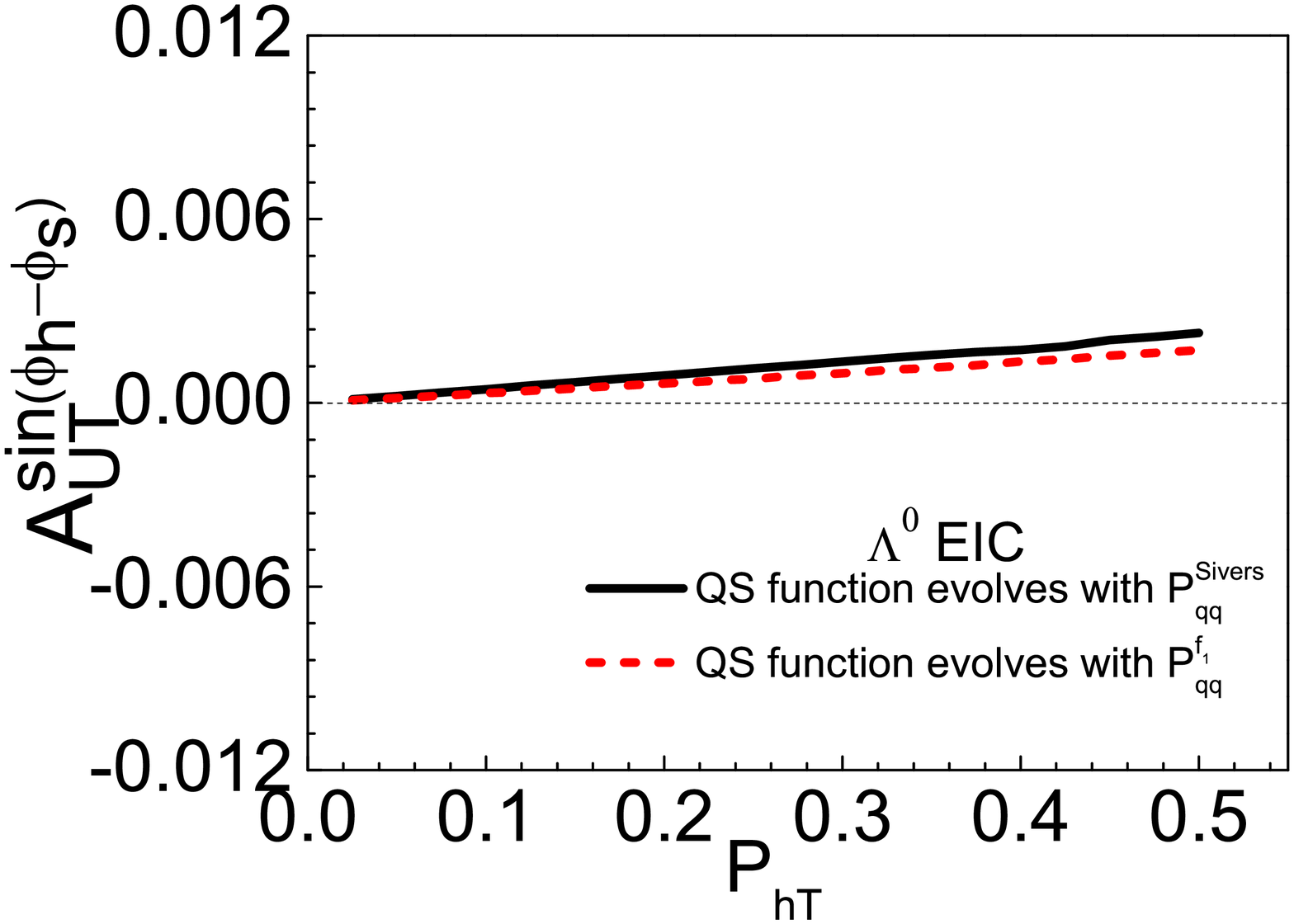}
	\includegraphics[width=0.3\columnwidth]{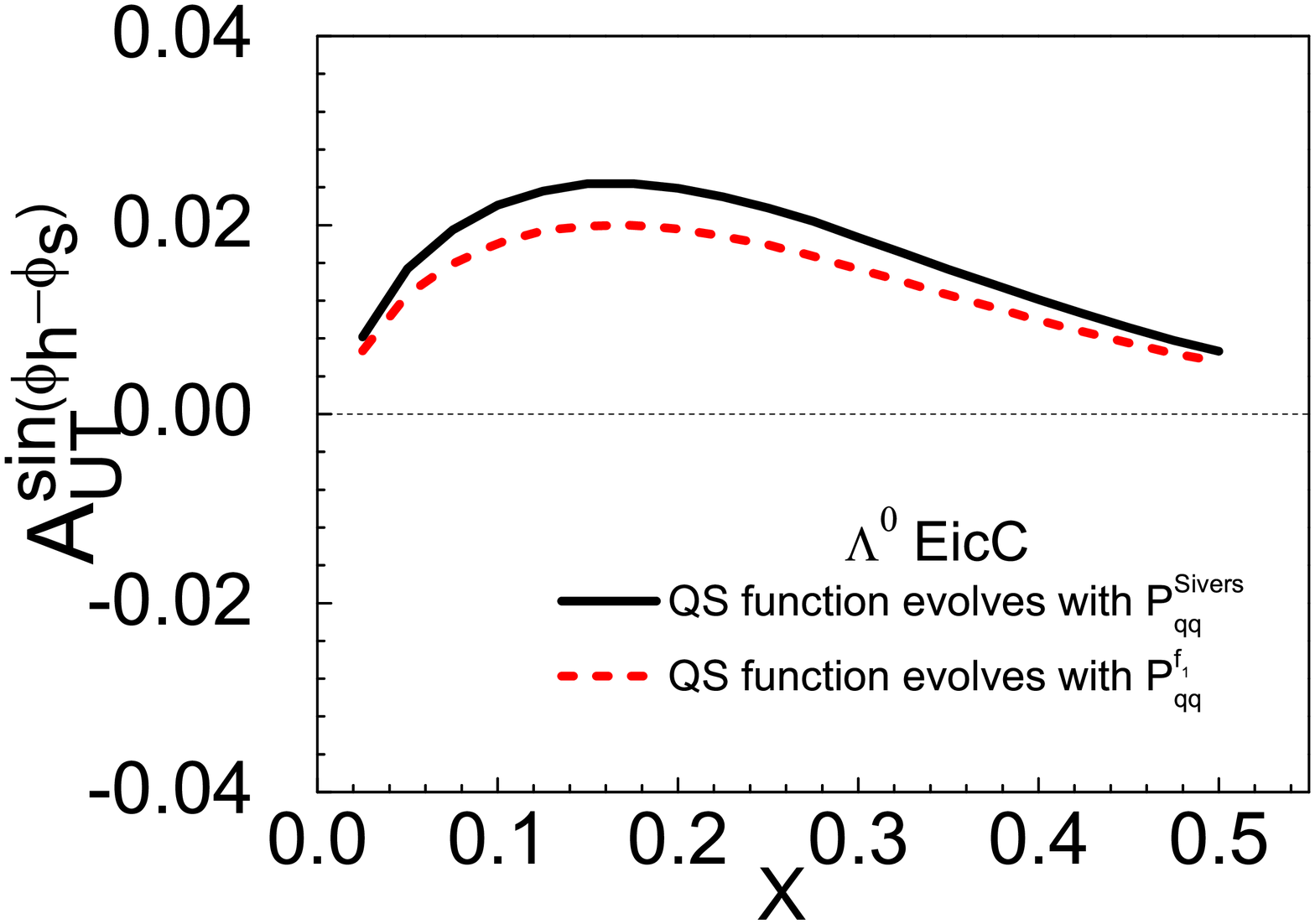}
	\includegraphics[width=0.3\columnwidth]{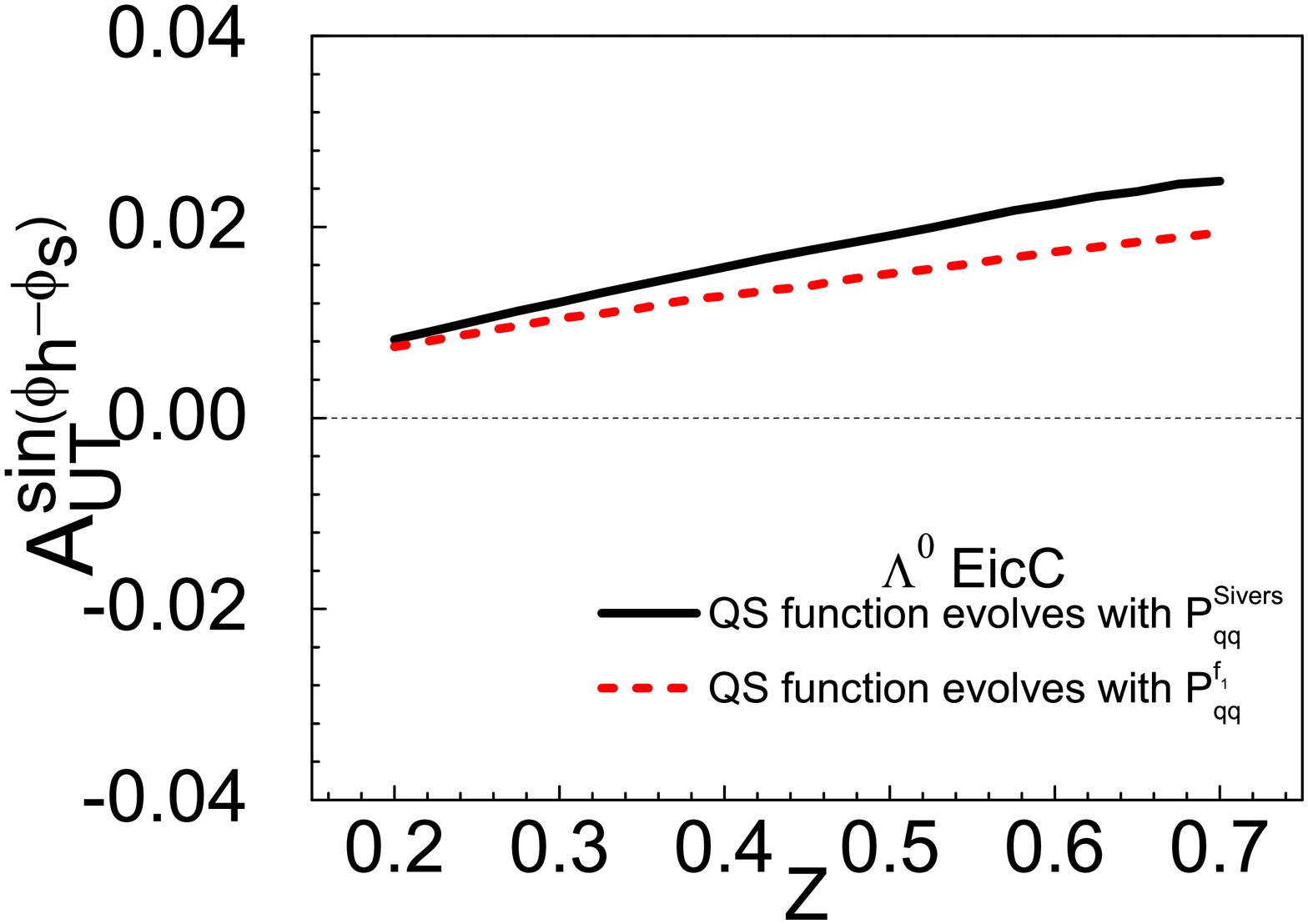}
	\includegraphics[width=0.3\columnwidth]{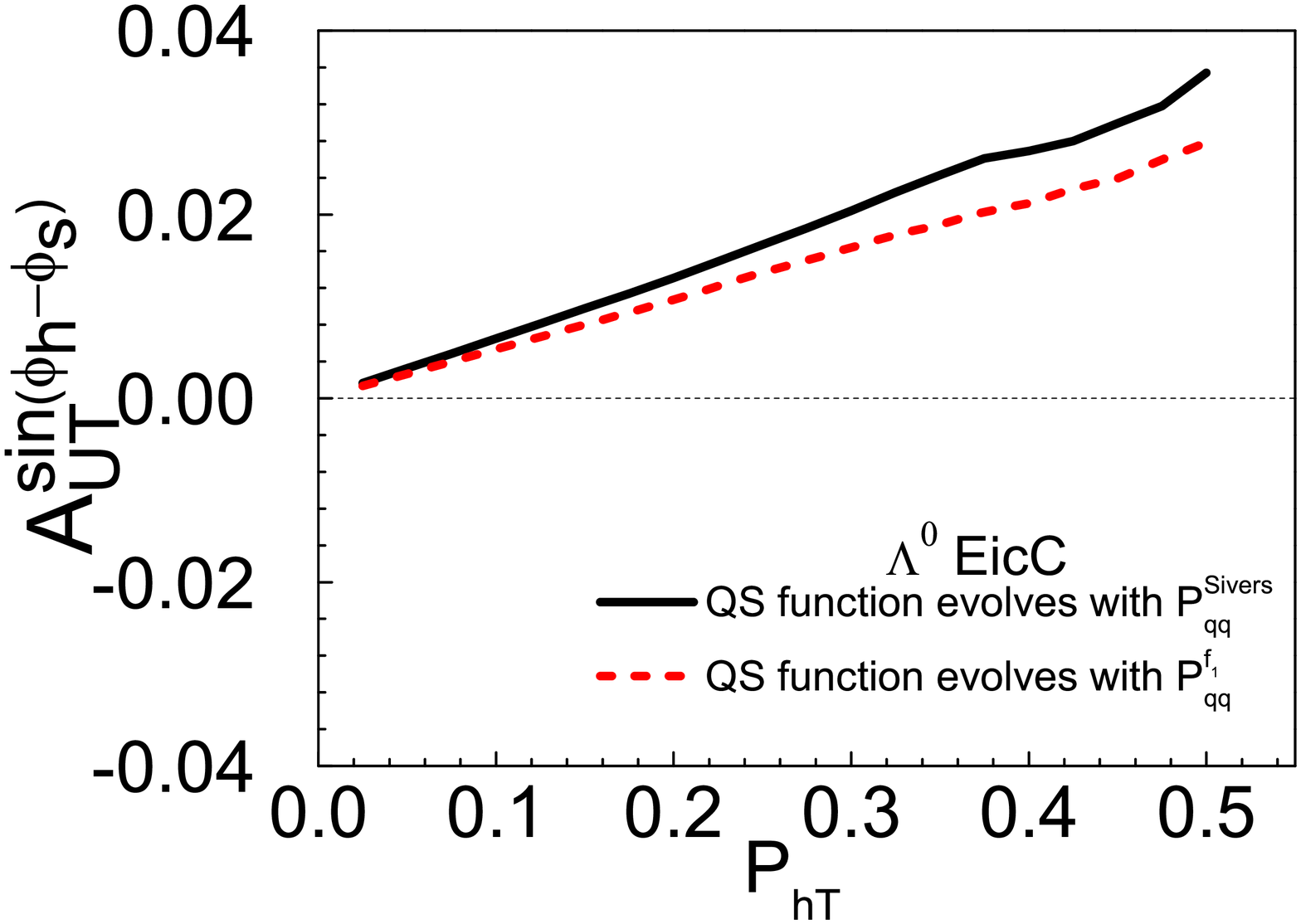}
	\caption{The Sivers asymmetry in semi-inclusive charged Kaon and $\Lambda$ hyperon produced SIDIS process at the kinematics of EIC and EicC as functions of $x$~(left panels), $z$~(middle panels), and $P_{hT}$~(right panels).}
	\label{fig:asy}
\end{figure}

The results are shown in Fig.~\ref{fig:asy}. The six rows in the figure depict the results of Sivers asymmetry of $K^{+}$ production, $K^{-}$ production, and  $\Lambda$ hyperon  production in the EIC and EicC kinematical region, respectively.
The left, central and right panels denote the Sivers asymmetry as the function of $x$, $z$ and $P_{hT}$, respectively.
In the figure, the black solid line represents the Sivers asymmetry that is obtained when the QS evolution kernel is included to evolve Eq.~(\ref{QS}).
The red dashed line represents the Sivers asymmetry obtained by using the evolution kernel for the unpolarized distribution function from QCDNUM evolution package to evolve Eq.~(\ref{QS}).

As can be seen from Fig.~\ref{fig:asy}, the Sivers asymmetries in the charged Kaon produced and in $\Lambda$ hyperon produced SIDIS process are sizable in the EIC and EicC kinematical region. Thus, the measurements of the Sivers asymmetry in these facilities provide an ideal tool to obtain the sea quark Sivers function as well as the flavor dependence of the Sivers function.
Also, it seems that the magnitude of the asymmetry in EicC is larger than that in EIC.
In addition, Fig.~\ref{fig:asy} shows that the value of the solid line is larger than the dashed line, which indicates that the Sivers asymmetry obtained using the evolution kernel of QS function is greater than that obtained using the unpolarized distribution function evolution kernel. One should note that the DGLAP evolution in the TMD effects may play some role in the future phenomenological analysis of the Sivers asymmetry.

We can also find from Fig.~\ref{fig:asy} that both the $z$-dependent and $P_{hT}$-dependent Sivers asymmetries in $K^{+}$ produced SIDIS process at the kinematical region of EIC and EicC are negative, and the magnitude of Sivers asymmetry increases with increasing $z$ or increasing $P_{hT}$.
While for $x$ dependent Sivers asymmetry of $K^{+}$ production in the EIC and EicC kinematical region, there is a node around $x=0.15$.
On the other hand, for the Sivers asymmetry of $K^{-}$  production, it is always positive in all cases. 
The Sivers asymmetry increases with $x$ at small $x$ region and decreases along $x$ at large $x$ region, and gradually increases with $z$ and $P_{hT}$.
In addition, by comparing the magnitudes of the Sivers asymmetry of $K^{+}$ produced with that of $K^{-}$ produced, we can see that the former one is significantly larger than the latter one, since for $K^{+}$ meson the constituent quarks  are $u$ and $\bar{s}$, and  for $K^{-}$ meson are $\bar{u}$ and $s$ quarks. Therefore there is relatively large contribution of valence $u$ quark Sivers function for $K^{+}$, while the contributions for $K^{-}$ are both from sea quark Sivers function.
It is known from Ref.~\cite{Echevarria:2014xaa} that the QS function of the valence quark is significantly larger than the that of the sea quarks, so the Sivers asymmetry in $K^{+}$ produced process is greater than that in $K^{-}$ produced process.
Finally, the Sivers asymmetry of $\Lambda$ hyperon production in both the EIC and EicC kinematics is positive, and the tendency of $x, z$ and $P_{hT}$ dependent asymmetry is generally consistent with that of $K^{-}$ meson.
In addition, the Sivers asymmetry in $\Lambda$ hyperon produced process is larger than that in charged Kaon $K^\pm$ produced process.

Therefore, the future higher precision EICs can provide a unique opportunity to extract the proton Sivers function of valence quark and sea quark from the charged Kaon $K^+$ production, to analyze the flavor dependence of the Sivers distribution function among sea quarks from $K^{-}$ produced SIDIS process, to investigate the flavor dependence from $\Lambda$ hyperon produced SIDIS process. Combining the experimental data from Drell-Yan process, it also provides an ideal tool to study the sign change of the Sivers function and the generalized universality of the T-odd distribution functions.

\section{CONCLUSION}
\label{sec:conciustion}

In this work, we apply the TMD factorization formalism to study the Sivers asymmetry with $\sin \left(\phi_{h}-\phi_{s}\right)$ modulation in charged $K^\pm$ produced and $\Lambda$ hyperon produced in SIDIS process at the kinematical configurations of EIC and EicC.
We take into account the TMD evolution effects of  distribution functions as well as fragmentation functions.
In our calculation we adopt the EIKV parametrization for the non-perturbative Sudakov-like form factor, the accuracy of the perturbative Sudakov-like form factor as well as the hard coefficients is kept at the NLL order.
Two different ways to deal with the energy dependence of Qiu-Sterman function associated with the Sivers function are applied.
The first approach is to assume the Qiu-Sterman function evolves as the same way as the unpolarized distribution functions $f_1(x,Q^2)$.
The second one is to evolve Qiu-Sterman function considering an approximate evolution kernel for the Qiu-Sterman function containing the homogenous terms by customising the DGLAP kernel.
The Sivers asymmetries are calculated as the functions of $x, z$, and $P_{hT}$. 
Our numerical results demonstrate that the Sivers asymmetries of charged Kaon production and $\Lambda$ hyperon production are measurable at the kinematics of EIC and EicC, with the magnitudes of around several percents.
The results show that the Sivers asymmetry in $K^\pm$ and $\Lambda$ hyperon production in SIDIS process can serve as an ideal tool to extract the information of sea quark Sivers function as well as to constrain the flavor dependence of Sivers function by utilizing future high energy and high luminosity EICs.
In addition, the difference between the Sivers asymmetries from considering two different evolution kernels suggests that the DGLAP evolution of the Qiu-Sterman function in the TMD evolution schemes will play a role in the phenomenological calculation, which should be considered in the future interpretation of experimental data as well as theoretical studies.

\section{ACKNOWLEDGMENTS}
This work is partially supported by the NSFC~(China) grants 11905187,11847217 and 12150013. X. Wang is supported by the China Postdoctoral Science Foundation under Grant No.~2018M640680.

\end{document}